\documentclass[aps,showpacs,amsmath,amssymb,pre,superscriptaddress,twocolumn,floatfix]{revtex4-1}

\usepackage{graphicx}
\usepackage{amsmath}
\usepackage{hyperref}
\usepackage{dcolumn}
\usepackage{bm}
\usepackage{color}

\usepackage[utf8]{inputenc}

\allowdisplaybreaks[1]

\begin{document}

\title{Local symmetry dynamics in one-dimensional aperiodic lattices}

\author{C.~Morfonios}
\affiliation{Zentrum f\"ur Optische Quantentechnologien, Universit\"{a}t Hamburg, Luruper Chaussee 149, 22761 Hamburg, Germany}

\author{P.~Schmelcher}
\affiliation{Zentrum f\"ur Optische Quantentechnologien, Universit\"{a}t Hamburg, Luruper Chaussee 149, 22761 Hamburg, Germany}
\affiliation{The Hamburg Centre for Ultrafast Imaging, Luruper Chaussee 149, 22761 Hamburg}

\author{P.~A.~Kalozoumis}
\affiliation{Department of Physics, University of Athens, GR-15771 Athens, Greece}

\author{F.~K.~Diakonos}
\affiliation{Department of Physics, University of Athens, GR-15771 Athens, Greece}

\date{\today}

\begin{abstract}
A unifying description of lattice potentials generated by aperiodic one-dimensional sequences is proposed in terms of their local reflection or parity symmetry properties. 
We demonstrate that the ranges and axes of local reflection symmetry possess characteristic distributional and dynamical properties which can be determined for every aperiodic binary lattice. 
A striking aspect of such a property is given by the return maps of sequential spacings of local symmetry axes, which typically traverse few-point symmetry orbits.
This local symmetry dynamics allows for a classification of inherently different aperiodic lattices according to fundamental symmetry principles.
Illustrating the local symmetry distributional and dynamical properties for several representative binary lattices, we further show that the renormalized axis spacing sequences follow precisely the particular type of underlying aperiodic order.
Our analysis thus reveals that the long-range order of aperiodic lattices is characterized in a compellingly simple way by its local symmetry dynamics.
\end{abstract}


\maketitle

\section{Introduction \label{intro}}

Aperiodic sequences of distinct potential units have long served as a flexible model for long-range order which goes beyond conventional crystalline periodicity.
As a means to approach disorder in an ordered manner, they pave the way for a fundamental characterization of condensed matter with respect to the combination of its structural and spectral properties \cite{Macia2006_RepProgPhys.69.397}.
A decisive aspect of order in one dimension is the presence of local reflection symmetries in a potential, that is, of distinct symmetric constituents that add up to a (typically) globally non-symmetric structure.
For infinitely extended systems, abundant local symmetries have been shown to lead to the absence of decaying eigenstates for the associated Schr\"odinger operator \cite{Hof1995_Comm.Math.Physics.174.149, Damanik2001_Ann.H.Poincare.2.927,Damanik2004_Proc.AMS.132.1957}.
For finite structures, {\it complete local symmetry}, that is, decomposability into symmetric parts, was recently shown to underly the occurrence--and enable the construction of--a class of perfectly transmitting resonances (PTRs) in quantum scattering \cite{Kalozoumis2013_PhysRevA.87.032113}.
A key ingredient that enables multiple PTRs in a single, globally non-symmetric potential, is the presence of nested local reflection symmetries on multiple scales.
Such a situation can be realized in one-dimensional (1D) aperiodic lattices that are decomposable into symmetric parts in different ways, and pertains to the case of classical wave scattering in nanophotonic devices \cite{Kalozoumis2013_phot} which are experimentally advantageous \cite{Macia2001_PhysRevB.63.205421,Negro2003_PhysRevLett.90.055501,Negro2012_Las.Phot.Rev.6.178,Poddubny2010_Physica.E.42.1871}.

Independently of their physical impact, local symmetries in 1D aperiodic sequences have been extensively studied under the name 'palindromes' in the combinatorics of words \cite{Droubay1995_Inf.Proc.Lett.55.217,Luca1997_Th.Comp.Sc.183.45,Droubay1999_Th.Comp.Sc.223.73,Allouche2003_Th.Comp.Sc.292.9,Borel2005_Th.Comp.Sc.340.334,Luca2005_Lect.Notes.Comp.Sc.3572.199,Glen2006_Th.Comp.Sc.352.31,Glen2009_Eur.J.Comb.30.510,Anisiu2006_PU.M.A.17.183}.
Here, the minimal reflection symmetric structural units are encoded as symbolic elements (the letters) of a set of given cardinality (the alphabet), out of which discrete lattices (the words) are constructed by concatenation.
Aperiodically ordered sequences are constructed by the iterative action of a given substitution (or inflation) rule on the set of lattice elements.
The presence of different palindromes (words that are read the same forwards and backwards) in a sequence is then expressed by its palindrome complexity function \cite{Allouche2003_Th.Comp.Sc.292.9}, which gives the number of contained palindromes of given length.
Rigorous mathematical results on the palindromicity of certain classes of (infinite) words have been obtained \cite{Allouche2003_Th.Comp.Sc.292.9,Droubay1999_Th.Comp.Sc.223.73,Damanik2000_Disc.Appl.Math.100.115,Luca2005_Lect.Notes.Comp.Sc.3572.199,Glen2006_Th.Comp.Sc.352.31}, an important part of which concerns palindromic prefixes (factors in the beginning of a word) \cite{Luca2005_Lect.Notes.Comp.Sc.3572.199,Luca2006_Int.J.Found.Comp.Sc.17.557,Fischler2006_J.Comb.Th.A113.1281}.
On the other hand, large effort has also been made for the computational determination of palindromes in arbitrary sequences \cite{Tomohiro2010_Lect.Notes.Comp.Sc.6393.135}, or even of gapped palindromes (having a non-symmetric central part) \cite{Kolpakov2009_Th.Comp.Sc.410.5365}, which are directly related to genome structure \cite{Lu2007_Funct.Int.Gen.7.221}.
Regarding complete local symmetry, mentioned above, the factorizability of finite words into (maximal length) palindromes has been demonstrated in closed form for binary words \cite{Ravsky2003_J.Aut.Lang.Comb.8.75}.

Aperiodic sequences are further studied in terms of the recurrence of factors or palindromes along given (classes of) words \cite{Glen2009_Eur.J.Comb.30.510,Allouche2003_Th.Comp.Sc.292.9,Masse2008_Proc.GASCOM.2008.53}, thereby giving a dynamical aspect to their structural properties.
In this context, an interesting question is whether--and to what extent--aperiodic sequences with arbitrary long-range order can be characterized by their {\it 'local symmetry dynamics'}, that is, the evolution of the {\it distances} between subsequent palindromes along a symbolic sequence.
Implementing such a description could have two important advantages: 
(i) allow for a classification which is independent of the exact symbolic composition of the palindromes in the considered sequence and 
(ii) reveal structural complexity at a deeper level where information can be encoded in a more compact way. 
As a by-product, the analysis of local symmetry dynamics could offer the possibility to introduce new families of symbolic sequences by modifying only the underlying local symmetry dynamics.

In this work, we address this task by considering the distribution of local reflection symmetries and the spacing of their axes in representative binary substitution sequences.
In contrast to the class-specific, rigorous combinatorial statements provided in the literature, we here perform numerical 'experiments' in order to shed light and to present a unifying viewpoint on the local symmetry properties of different classes of aperiodic lattices.
The analysis reveals that the long-range order and complexity of well-known substitution sequences can be simply encapsulated within their local symmetry dynamics.
In particular, we show that the return maps of palindrome spacings for given palindrome lengths consist of finite sets of strongly collinear points, with real-space trajectories which follow the original substitution rule.
Apart from compact characterization, this allows for an immediate recognition of qualitative--but also quantitative--common features and differences among various setups.

The paper is organized as follows.
In Sec.~\ref{notation} the basic concepts of symbolic aperiodic sequences are introduced and defined, along with a notation suitable for our needs.
In Sec.~\ref{lsdd} the local symmetry properties of representative aperiodic sequences are individually addressed and analyzed.
In particular, the different symmetry aspects--including its distribution, dynamics, and scaling thereof--are first introduced for the standard Fibonacci lattice, and then examined for generalized Fibonacci, period-doubling and Thue-Morse sequences, which are finally contrasted with the structurally different Cantor lattice.
Section \ref{conclusion} summarizes and concludes the present work.

\section{Basic concepts and notation \label{notation}}

Before starting the local symmetry analysis of aperiodic sequences, we briefly introduce here the global notation and terminology appropriate for our purposes.
Since we will consider, throughout this work, 1D potentials that are composed of mirror symmetric building blocks (scatterers), we can make use of the basic definitions for symbolic sequences.
Different scatterers are thus symbolized by different {\it letters} $a_j$, $j = 1,2,...,N$, which are elements of a finite {\it alphabet} $\mathcal A$ of cardinality $|\mathcal A| = N$.
A finite 1D lattice of scatterers thus corresponds to a combination of letters concatenated into a {\it word} 
\begin{equation}w = x_1x_2\cdots x_n, ~~ x_i \in \mathcal{A}, 
\end{equation} 
of length $l_w \equiv |w| = n$.
The set of all such words defines the {\it language} $\mathcal L = \mathcal A^*$, which is the so called free monoid generated by $\mathcal A$, including the empty word $\epsilon$ of length zero (the identity element of $\mathcal L$).
Within $\mathcal L$, words can thus also be concatenated to from new words.
A {\it factor} $f$ of $w$, corresponding to a number of consecutive scatterers in the lattice, is a word such that $w = ufv$ with $u,v \in \mathcal L$.
If $u = \epsilon$ ($v = \epsilon$), then $f$ is in the beginning (end) of the sequence and is called {\it prefix} ({\it suffix}) of $w$.

The reversal of the word $w$ above is defined as 
\begin{equation}\tilde w = x_nx_{n-1}\cdots x_1. 
\end{equation}
If $\tilde w = w$, then $w$ is called {\it palindrome}, i.e., a word which reads the same forwards and backwards.
Palindromes thus correspond to reflection symmetric sequences of scatterers, and palindromic factors to {\it locally symmetric} parts of a potential.
We denote by $\mathcal{P}(w) \subset \mathcal{L}$ the set of all palindromic factors of a (finite or infinite) word $w$, and by $\mathcal{P}_l(w)$ its subset of palindromic factors of length $l$.
The cardinality of $\mathcal{P}_l(w)$ is the {\it palindrome complexity function}, $p_w(l)$, which thus gives the number of {\it different} palindromes of given length $l$ contained in $w$, i.e., irrespectively of their position.
The palindromic factors of generic binary words have been extensively studied in terms of the associated palindrome complexity function \cite{Allouche2003_Th.Comp.Sc.292.9,Brlek2004_Int.J.Found.Comput.Sci.15.293}, but also in relation to palindromic factorization \cite{Ravsky2003_J.Aut.Lang.Comb.8.75,Frid2012_TUCS.1063,Wen1994_Eur.J.Comb.15.587,Glen2006_Th.Comp.Sc.352.31,Chuan2012_Th.Comp.Sc.440.39}, i.e. decomposition into locally symmetric parts.

\begin{figure}[t]
\includegraphics[width=\columnwidth]{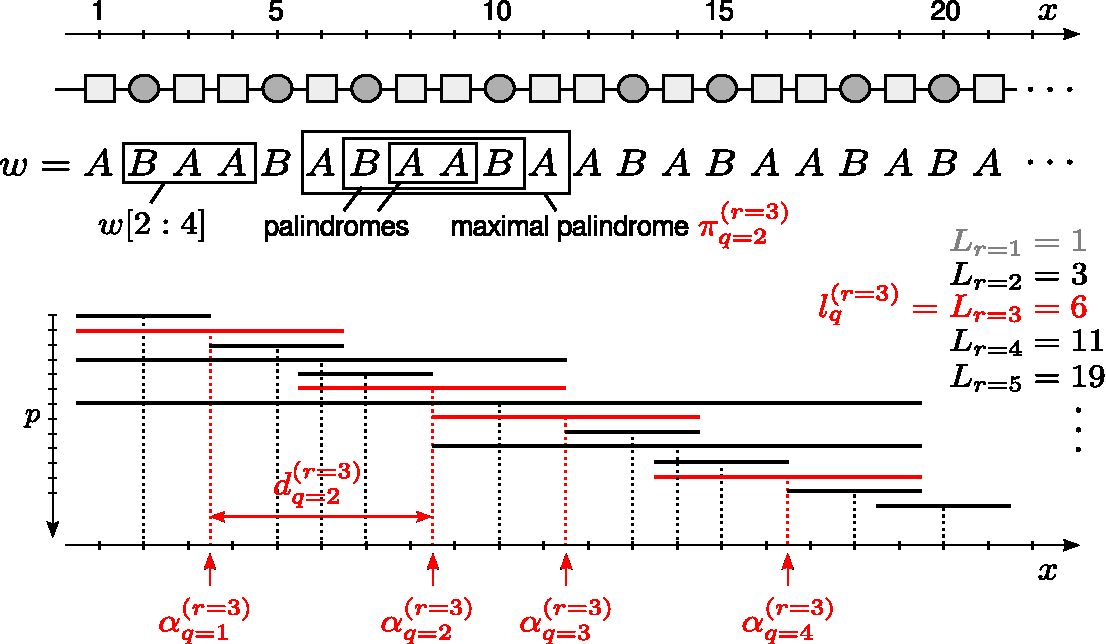}\vspace{.5cm}
\caption{\label{fig.sketch} (Color online) An aperiodic 1D lattice of two types of equidistant, arbitrary reflection symmetric scatterers (depicted as squares and circles) is mapped to a binary symbolic sequence of letters ($A$ and $B$, respectively), constituting the word $w$ (whose shown part coincides with the $4$-th generation Fibonacci word, see text).
The letter symbols are set to unit length and centered at the positive integers $x=i \in \mathbb Z^+$ on the $x$-axis. 
Maximal palindromic factors of $w$ with more than one letters, denoted as $w[i:j]$, are depicted as horizontal line segments of length $l_p$ centered at $\alpha_p$, counted by the index $p$ in order of increasing $\alpha_p$.
The index $q = 1,2,...$ counts the maximal palindromes of common length $l^{(r)}_q = L_r$ occurring in $w$, again in order of increasing position $\alpha^{(r)}_q$, as shown (in red) for the length $L_3 = 6$.
Single-letter palindromes (with $L_{r=1}=1$, in this case $A$) are not shown.}
\end{figure}

With the symbolic correspondence between a given 1D completely locally symmetric potential and a word $w$, we map the symmetric scatterers to the letters $x_i$ (of unit length) of $w$ centered at the positive integers $x = i \in \mathbb Z^+$ on the $x$-axis, and denote by $w[i:j]$ the factor $x_i x_{i+1} \cdots x_{j}$ of $w$ (see Fig.~\ref{fig.sketch}).
Having introduced this coordinate system, any (palindromic) factor $\pi = w[i:j]$ is represented as $(\alpha, l)$, where
\begin{equation}
 \alpha = \frac{i + j}{2}, ~~~l = |\pi| = |j - i| + 1
\end{equation}
are its center and length, respectively (that is, the reflection axis position and the range of this locally symmetric part in the lattice).

The palindrome $\pi$ is called {\it maximal} if it is the largest palindrome centered at $\alpha$, i.e., if any palindromic factor of $w$ centered at $\alpha$ has length $l' \leqslant l$.
We denote by $\mathcal{M}^x(w)$ the set of all pairs $(\alpha_p,l_p)$, $p=1,2,...$, of coordinates of maximal palindromes $\pi_p$ with length $l_p \geqslant 2$, ordered in increasing $\alpha_p$ along the $x$-axis.
For each occurring maximal palindrome length $L_r$, ($r = 1,2,...$ with $L_{r+1} > L_r$), we define the subsets 
\begin{equation} 
\mathcal{M}^x_{L_r}(w) = \{(\alpha_p,l_p) \equiv (\alpha^{(r)}_q,l^{(r)}_q) \in \mathcal{M}^x(w)~|~l_p = L_r \}, 
\end{equation}
where the index $q = 1,2,...$ counts the maximal palindromes of given length $L_r$, again ordered in their position $\alpha^{(r)}_q$ (see Fig.~\ref{fig.sketch}).
Mapping the elements of $\mathcal{M}^x_{L_r}(w)$ (i.e., word coordinates) back to the corresponding words, 
\begin{equation}
(\alpha^{(r)}_q, l^{(r)}_q) \to \pi^{(r)}_q \in \mathcal{L},
\end{equation} 
we obtain the set ${\mathcal{M}}_{L_r}(w)$ of different maximal palindromic factors of length $L_r$, with $\cup_r {\mathcal{M}}_{L_r}(w) \equiv {\mathcal{M}}(w)$.

The construction of a lattice with aperiodic order can be simply achieved by starting from one (or more) letter(s) and then iteratively apply a single rule which substitutes letters (or words) with (larger) words, thereby ensuring the presence of long-range order among the scatterers of the emerging potential.
Such a substitution (or {\it inflation}) rule $\sigma$ is formally defined as a map $\sigma : \mathcal L \to \mathcal L$, supplied with the property 
\begin{equation}\sigma(uv) = \sigma(u)\sigma(v), ~~u,v \in \mathcal L,\end{equation} 
as well as 
\begin{equation}\sigma^k(u) = \sigma(\sigma^{k-1}(u)),~~~  {\rm with}~~ \sigma^0(u) \equiv u.
\end{equation}
In particular, its action on the word $w = x_1x_2\cdots x_n$, $x_i \in \mathcal{A}$, simply becomes 
\begin{equation}
 \sigma(w) = \sigma(x_1)\sigma(x_2)\cdots \sigma(x_n).
\end{equation}
In the following, we will consider lattices corresponding to words on a two-letter (binary) alphabet $\mathcal{A} = \{A,B\}$, constructed through an inflation rule $\sigma$ defined by its action on the single letters, using the shorthand notation 
\begin{equation}\sigma(A,B) \equiv (\sigma(A),\sigma(B)).\end{equation}

\section{Local symmetry distribution and dynamics \label{lsdd}}

To introduce the concept of {\it local symmetry dynamics}, which is the main task of this section, we focus on the space of maximal palindromes, that is, the maximal locally symmetric parts of the corresponding scatterer lattice. 
As a first step we classify the maximal palindromes, as they appear in the infinite word of the considered symbolic sequence, with respect to their length $L_r$, where $r$ counts the possible different length values in increasing order: $L_r < L_{r+1}$.
Then, for a given $L_r$, we determine the positions $\alpha_q^{(r)}$ of the symmetry axes of the maximal palindromes with this length (i.e., those belonging to the set $\mathcal{M}^x_{L_r}(w)$), and represent their distribution through the plot $L_r$ vs. $\alpha_q^{(r)}$ for $r=1,2,..$.
The index $q$ counts the maximal palindromes of given length $L_r$ for increasing symmetry axis position $\alpha_q^{(r)}$.
This local symmetry distribution characterizes the palindrome composition of the considered symbolic sequences in a symbol-independent manner, and can be used for their classification with respect to structural features.
The local symmetry dynamics is then determined for each $r$ by the change of the distance between subsequent maximal palindrome symmetry axes, $d_q^{(r)}=\alpha_q^{(r)} - \alpha_{q-1}^{(r)}$, as $q$ increases. 
We choose here to present this dynamics through the return map $d_q^{(r)}$ vs. $d_{q-1}^{(r)}$, which turns out to be an excellent tool to reveal dynamical scaling properties as well as correlations at different scales.
The latter are, as will be explained below, directly related to the fact that the considered symbolic sequences possess nested local symmetries, that is, systematically and multiply overlapping symmetric parts of the lattice.
The detailed local symmetry dynamics, which will be shown to follow the corresponding original inflation rule of a given sequence, is finally illustrated by explicit $d_q^{(r)}$-trajectories for selected $r$.
In the following we will present results of this analysis applied to five different types of aperiodic sequences: (A) Fibonacci, (B) generalized Fibonacci, (C) period doubling, (D) Thue-Morse and (E) Cantor lattices.

\subsection{Fibonacci sequence: a prototype ordered aperiodic lattice \label{fib}}

Let us first analyze the local symmetries of a lattice composed according to the famous Fibonacci sequence, which constitutes a prominent example of quasiperiodic order \cite{Macia2001_PhysRevB.63.205421,Wang2000_PhysRevB.62.14020}.
As a symbolic two-letter sequence, the $k$-th order Fibonacci word is generated by iterative application of the inflation rule 
\begin{equation}\sigma_F(A,B) = (AB,A),
\end{equation} 
conventionally starting with the letter $A$: 
\begin{equation}w_F^{(k)} = \sigma_F^k(A), ~~k=0,1,2,...
\label{eq.fib}
\end{equation}
For $k \geqslant 2$, it has the recursion property 
\begin{equation}w_F^{(k)}=w_F^{(k-1)}w_F^{(k-2)},\end{equation}
so that its length $|w_F^{(k)}|=|w_F^{(k-1)}|+|w_F^{(k-2)}|$ follows the numeric Fibonacci sequence $1,2,3,5,8,13,...$, with 
\begin{equation}{\lim_{k\to\infty}}\frac{|w_F^{(k)}|}{|w_F^{(k-1)}|} = \frac{1+\sqrt{5}}{2} = \varphi_g,\end{equation}
the so-called {\it golden} mean.
Denoting by $w_F \equiv \lim_{k\to\infty}w_F^{(k)}$ the infinite Fibonacci word, we note that any sufficiently long factor of $w_F$ (i.e., not necessarily a $k$-th order prefix $w_F^{(k)}$) is characterized by the aperiodic order induced by the inflation rule.

\begin{figure}[t]
\includegraphics[width=.88\columnwidth]{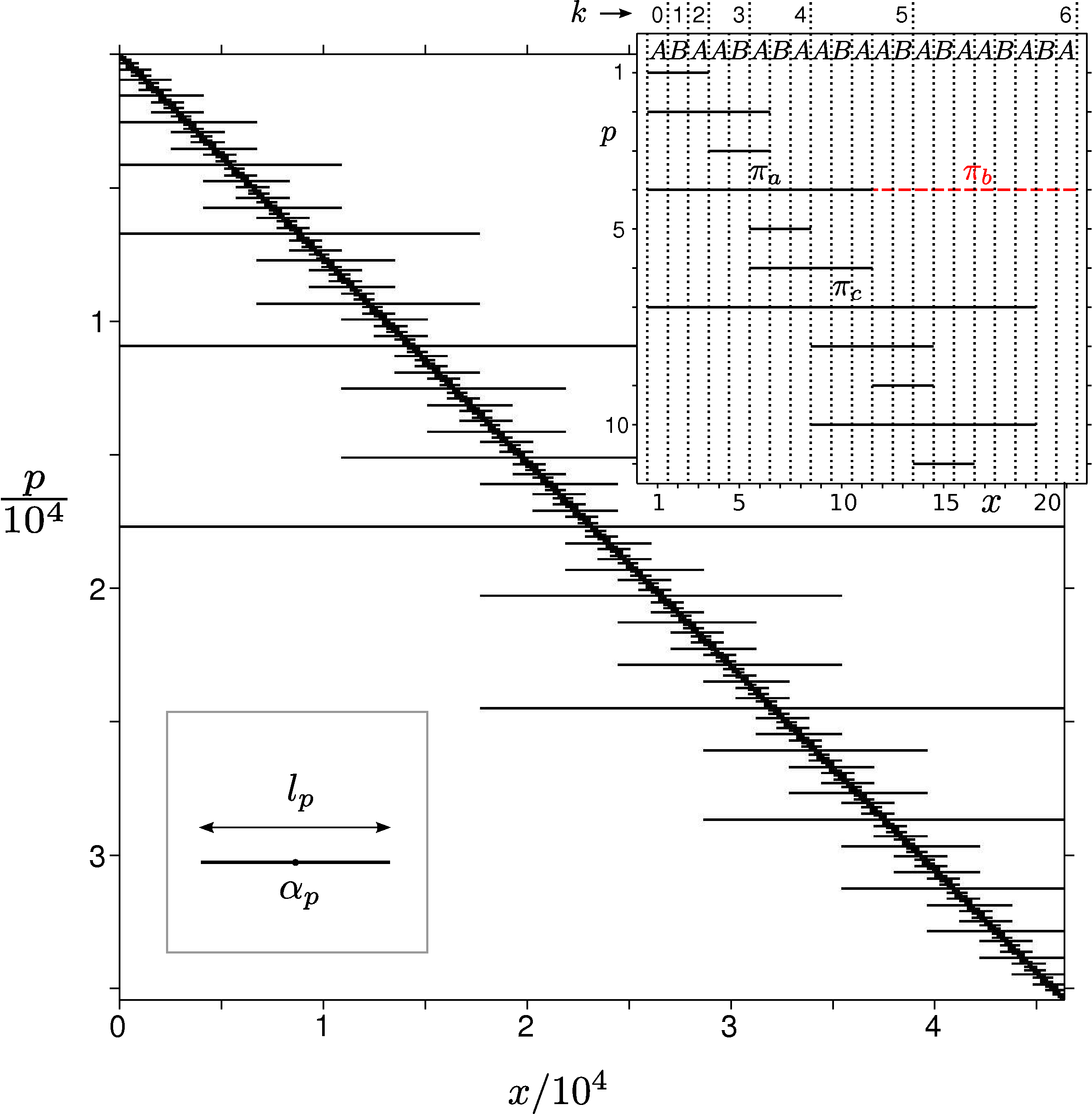}\vspace{.5cm}
\caption{\label{fig.fib-pal} (Color online) Local symmetries of the Fibonacci lattice, corresponding to maximal palindromes of length $l_p$ in the letter sequence $w_F$, represented as line segments in order $p$ of their occurrence along the $x$-axis (see coordinates defined in Fig.~\ref{fig.sketch}). Those maximal palindromes of lengths $\geqslant 1$ are shown, which are contained within the $22$-th order Fibonacci word $w_F^{(22)}=\sigma_F^{22}(A)$. The inset shows the $6$-th generation word $w^{(6)}_F$ (the generations are counted by the index $k$), containing the first four prefix palindromes with $l > 1$. $w^{(6)}_F$ can be decomposed into a prefix maximal palindrome $\pi_a$ and a non-maximal palindrome $\pi_b$ (distinguished by a the dashed line), while its last prefix palindrome $\pi_c$ reaches up to its two last letters $BA$ (see text).}
\end{figure}

The local reflection symmetries of the Fibonacci lattice are illustrated in Fig.~\ref{fig.fib-pal}, where its maximal palindrome coordinates $(\alpha_p, l_p)$, with $p$ ordered in increasing $\alpha_p$, are shown (excluding the single-letter palindromes).
The infinite Fibonacci word $w_F$ belongs to the class of {\it Sturmian} words \cite{Luca1997_Th.Comp.Sc.183.45,AlloucheShallit2003}, which can in fact be defined as the (infinite) binary words with palindromic complexity  \cite{Droubay1995_Inf.Proc.Lett.55.217,Droubay1999_Th.Comp.Sc.223.73,Allouche2003_Th.Comp.Sc.292.9}
\begin{equation}p_w(l={\rm even, odd}) = 1,2.
\end{equation}
Note, however, that the number of different {\it maximal} palindromes of given length $l$ is generally {\it not} given by $p_w(l)$.
E.g., for $l = 3$, $\mathcal{P}_3 = \{ABA,BAB\}$, but ${\mathcal{M}}_3 = \{ABA\}$, since $BB$ is not a factor of $w_F$ and therefore $BAB$ cannot be maximal.

Further, Sturmian aperiodicity can be geometrically connected to irrational numbers (like $\varphi_g$) by the so-called 'cut and project' scheme \cite{Baake1999_Lett.Math.Phys.49.217,Moody1999,Guimond2003_J.Th.Nombr.Bord.15.697,Poddubny2010_Physica.E.42.1871}:
they are constructed from the projection of the vertices of a square two-dimensional lattice, lying within a stripe of given width \cite{Poddubny2010_Physica.E.42.1871} and with irrational slope, onto the direction of the stripe.
The segments between the projected vertices (being of two possible lengths) are mapped to letters $A$ and $B$, thus constituting a word.
The slope of the stripe is also called the slope of the word; if it equals $1/\varphi_g$, one obtains $w_F$.

Regarding their decomposability, finite Sturmian words can be uniquely written as \cite{Luca1995_Inf.Proc.Lett.54.307,Luca1997_Th.Comp.Sc.183.45}
\begin{equation}w = \pi_a \pi_b = \pi_c s, \label{eq.sturm-dec}
\end{equation}
where $\pi_a, \pi_b, \pi_c$ are palindromes and $s = AB$ or $BA$.

For the Fibonacci word generations $w^{(k)}_F$, we have $s = AB$ ($BA$) for even (odd) $k$. 
This can be seen in the inset of Fig.~\ref{fig.fib-pal} for the generations $k = 3,4,5,6$, with corresponding prefix palindrome lengths $|\pi_c| = 3, 6, 11, 19$.
As also seen, the palindromic decomposition can in general not be achieved solely with maximal palindromes.
E.g., for $k=6$, the palindrome prefix with $|\pi_a| = 11$ is maximal, but concatenated to a non-maximal palindrome with $|\pi_b| = 10$.

Further, any sufficiently large prefix of the Fibonacci sequence can be factorized in multiple ways into palindromes of different lengths \cite{Wen1994_Eur.J.Comb.15.587,Ravsky2003_J.Aut.Lang.Comb.8.75,Glen2006_Th.Comp.Sc.352.31,Frid2012_TUCS.1063}.
Therefore, any finite scatterer lattice with Fibonacci aperiodic order is {\it completely locally symmetric} (completely decomposable into locally symmetric parts) at {\it multiple} scales.
This property relies on the distribution of palindromes of each given length along the sequence, determining which palindromes can be successively concatenated to construct a given finite lattice.

\begin{figure}[t]
\includegraphics[width=.88\columnwidth]{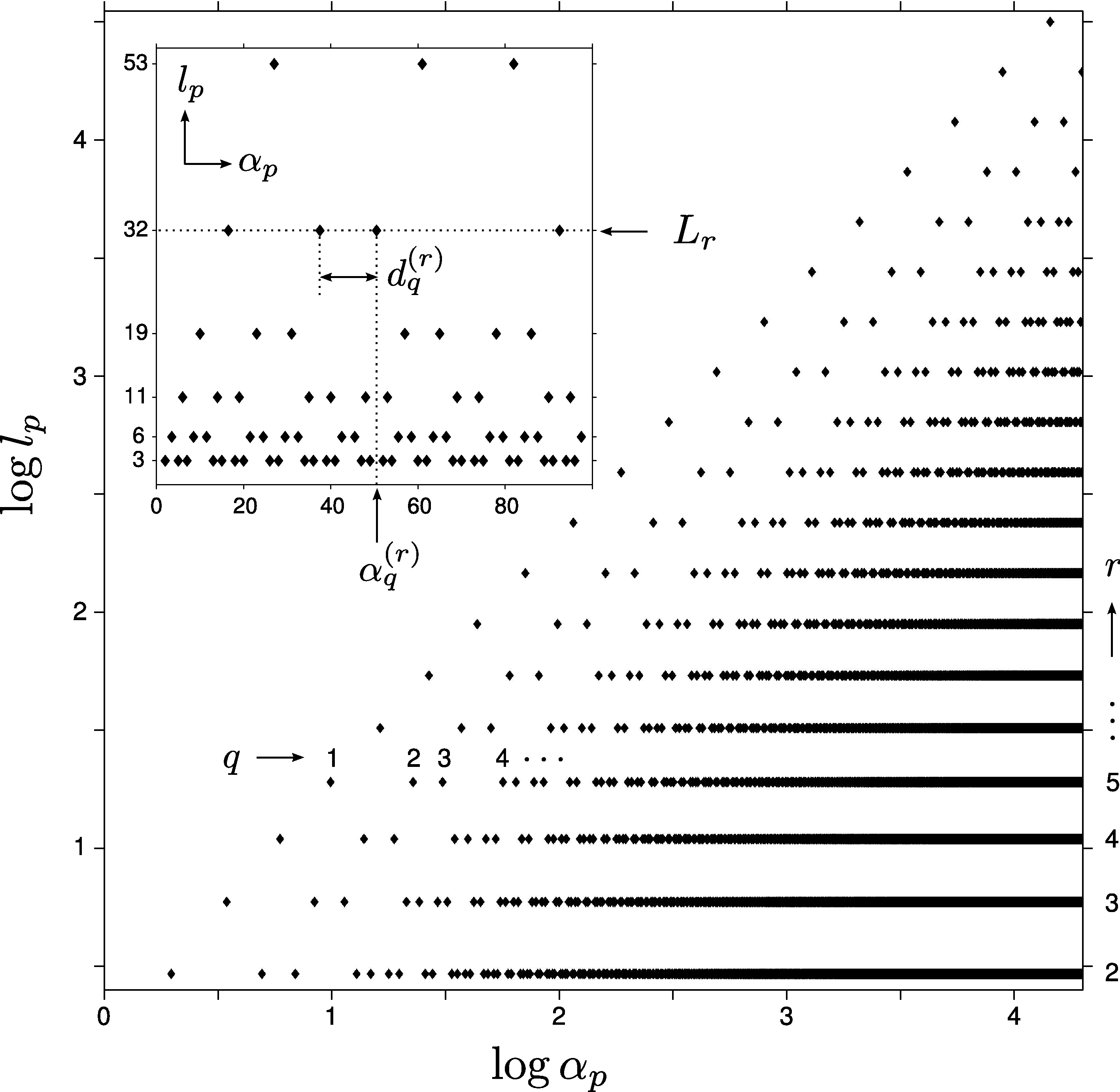}\vspace{.5cm}
\caption{\label{fig.fib-pal-cl} Local symmetry distribution of the Fibonacci lattice. Pairs $(\alpha_p,l_p)$ of axis positions and lengths of maximal palindromes $\pi_p$, are plotted on base-$10$ logarithmic scale. For each occurring symmetry range $L_r$ ($r=1,2,...$, $L_1$ not plotted), the index $q$ counts the maximal palindromes $\pi^{(r)}_q \in \mathcal{M}_r$ with $l^{(r)}_q = L_r$, ordered in increasing axis position $\alpha^{(r)}_q$ (horizontally collinear points). The inset shows the linear distribution of local symmetry axes for the first occurring ranges.}
\end{figure}

To analyze the local symmetry distribution of the Fibonacci lattice, we plot in Fig.~\ref{fig.fib-pal-cl} the elements of 	$\mathcal{M}^x(w_F)$, i.e., the length $l_p$ of each maximal palindrome versus its axis position $\alpha_p$ (excluding pairs corresponding to single letter maximal palindromes).
As we see, only certain lengths occur, which constitute the sequence $(L_r)_{r \geqslant 1} = 1,3,6,11,19,32,53,...$ (see inset of Fig.~\ref{fig.fib-pal-cl}).
A remarkable feature of the sets $\mathcal{M}^x_r(w_F)$ (horizontally collinear points in Fig.~\ref{fig.fib-pal-cl}) is that their first element $(\alpha^{(r)}_{q=1},l^{(r)}_{q=1})$ always corresponds to a {\it prefix} of $w_F$, so that 
\begin{equation}
\alpha^{(r)}_{q=1} = \frac{L_r+1}{2}.
\label{eq.prefix}
\end{equation}
In particular, for the Fibonacci sequence we thus have
\begin{equation}
 \alpha^{(r)}_{q=1} = \frac{|w_F^{(r)}|-1}{2}, ~~~~ l^{(r)}_{q=1} = L_r =  |w_F^{(r)}|-2,
\end{equation}
corresponding to the palindrome $\pi_c$ in $w_F = \pi_c s$, as defined above in Eq~(\ref{eq.sturm-dec}), for each word generation $k=r$ \cite{Luca1997_Th.Comp.Sc.183.45}.
This means that, at these special axis positions arbitrarily deep within the lattice, there occur local symmetries which extend to the beginning (left boundary) of it, which we coin {\it complete local symmetries}.
In view of wave scattering, such complete local symmetries are important, because they can render the lattice transparent \cite{Kalozoumis2013_PhysRevA.87.032113} over a large part from its beginning, despite the complex, aperiodic character of the underlying medium.
In particular for the Fibonacci lattice, the trailing two scatterers ($s=AB$ or $BA$ in Eq.~(\ref{eq.sturm-dec})) could be discarded from any Fibonacci word $w^{(k)}_F$, in order to produce perfectly transmitting resonances \cite{Kalozoumis2013_PhysRevA.87.032113,Kalozoumis2013_phot} in a globally symmetric system with quasiperiodic order.

As anticipated from the above, the ratio of successive palindromic lengths in $w_F$ converges to the golden mean, 
\begin{equation}
\lambda \equiv \lim_{r\to\infty}\frac{L_{r+1}}{L_r} = \varphi_g.
\label{eq.lambda}
\end{equation}
In fact, the Fibonacci word $w_F$ is a special case of Sturmian words with {\it abundant} palindromic prefixes, studied in Ref.~\cite{Fischler2006_J.Comb.Th.A113.1281}:
it is the one with the smallest ratio $\lambda$ (apart from periodic words with $\lambda = 1$), thus being the most dense in prefix palindromes.
Furthermore, the fact that each prefix palindrome is centered within the previous one (since $\varphi_g < 2$), suggests that $w_F$ can be {\it constructed} in terms of its palindromic prefixes, as is indeed the general case for this class of words \cite{Fischler2006_J.Comb.Th.A113.1281}.

\begin{figure}[t!]
\includegraphics[width=.88\columnwidth]{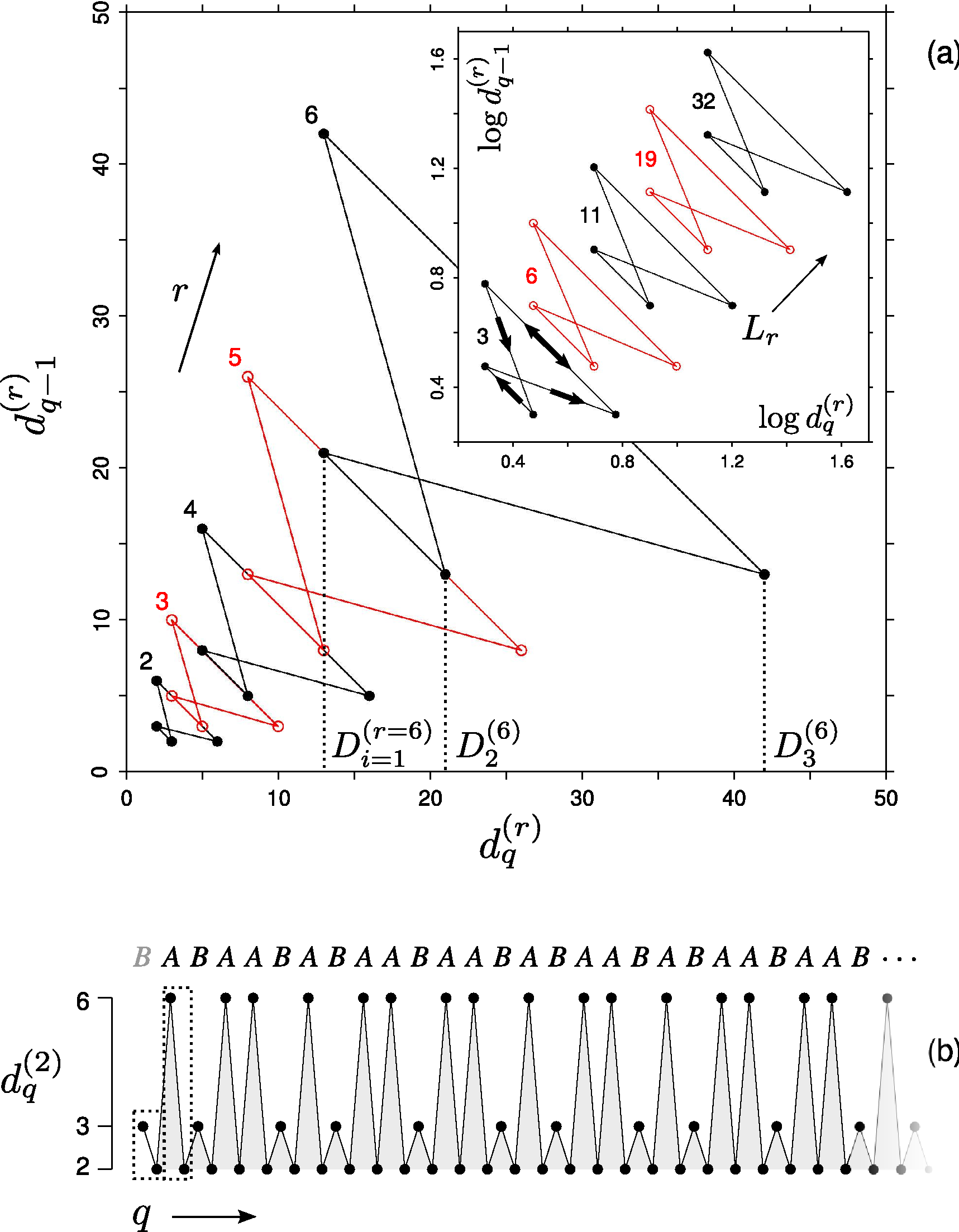}\vspace{.5cm}
\caption{\label{fig.fib-pal-spacing} (Color online) (a) Local symmetry spacing return maps for the Fibonacci lattice. For each occurring symmetry range $L_r$, each distance $d^{(r)}_q = \alpha^{(r)}_q - \alpha^{(r)}_{q-1}$ between axes of consecutive palindromes is plotted versus the previous distance $d^{(r)}_{q-1}$, with full (empty) circles for even (odd) $r$. 
Each map consists of three possible spacings $D^{(r)}_i$ ($i=1,2,3$), and consecutive map points are connected by lines. 
Note that the plot consists of multiple disconnected maps plotted together, and distinguished by the alternately filled/empty circles.
The inset repeats the plot on base-$10$ logarithmic scale, with maps labeled by the corresponding $L_r$; thick arrows on the lowest map indicate possible direction of traversal of any map.
(b) Local symmetry spacing trajectory (i.e., the $d^{(r)}_q$ plotted in order of occurrence $q$ for fixed $r$) for range $L_{r=2} = 3$ in the Fibonacci lattice. By mapping the pairs $D^{(2)}_3D^{(2)}_1$ and $D^{(2)}_2D^{(2)}_2$ of consecutive spacings (marked by dotted boxes) to letters $A$ and $B$, respectively, the trajectory reproduces the original sequence (see text).}
\end{figure}

Let us now turn to the local symmetry dynamics of the Fibonacci lattice, that is, the evolution of consecutive {\it spacings} between maximal palindromes of given length along the word $w_F$.
Fig.~\ref{fig.fib-pal-spacing}(a) shows, for each length $L_r$, the return map of the spacings 
\begin{equation}d^{(r)}_q \equiv \alpha^{(r)}_q - \alpha^{(r)}_{q-1}~~~(q \geqslant 2)
\end{equation}
between consecutive local symmetry axes ($d^{(r)}_q$ is also indicated in Figs.~\ref{fig.sketch} and \ref{fig.fib-pal-cl}).
The $d^{(r)}_q$ for each $L_r$ take on only three possible values
\begin{equation}D^{(r)}_i ~~(i=1,2,3),~~{\rm with}~~D^{(r)}_1 < D^{(r)}_2 = \frac{D^{(r)}_3}{2},
\end{equation}
in a succession which forms a four-point orbit in each return map.
We see that, for any given symmetry range $L_r$, none of the spacings $D^{(r)}_i$ is repeated consecutively as $q$ increases ($d^{(r)}_q \neq d^{(r)}_{q-1}$), and that $d^{(r)}_q$ returns to the minimal spacing $D^{(r)}_1$ for every second $q$.

We further observe that the spacing return maps of consecutive local symmetry ranges are correlated in the following two-fold {\it nested} sense:
Firstly, the middle spacing for each palindrome length coincides with the minimal spacing for the next length, 
\begin{equation}D^{(r)}_2 = D^{(r+1)}_1.\end{equation}
Secondly, we have that
\begin{equation}D^{(r)}_1 + D^{(r)}_3 = D^{(r+1)}_1 + D^{(r+1)}_2,\end{equation}
so that consecutive return maps have four collinear points along the lines with slope $-1$, as seen in the linear plot of Fig.~\ref{fig.fib-pal-spacing}(a).

With respect to the spacing return maps, there are here two properties which demonstrate how the aperiodic order of the Fibonacci sequence is encoded in its local symmetry dynamics:\\
(i) The sequence of minimal spacings coincides with the numeric Fibonacci sequence, 
\begin{equation}D^{(r)}_1 = |w^{(k=r)}_F| ~~(= 1,2,3,5,8,...),
\end{equation}
counting here also the single-letter maximal palindromes $A$ (not plotted).\\
(ii) The spacing return maps for different $L_r$ present a characteristic scaling, which converges \cite{lim} to the inverse slope of $w_F$, 
\begin{equation}
\lim_{r\to\infty}\frac{D^{(r)}_2}{D^{(r)}_1} = \varphi_g.
\label{eq.fib-spacing}
\end{equation}

The quasiperiodicity of the Fibonacci lattice is, finally, manifest also in the individual {\it trajectories} of the axis spacings for given symmetry ranges, i.e. the order in which the points of the return map orbits are traversed (this information is obviously not available from the return maps alone).
The spacing trajectory for $L_{r=2} = 3$ is shown in Fig.~\ref{fig.fib-pal-spacing}(b).
We see that the sequence of spacings for odd $q$, with elements $D^{(2)}_3, D^{(2)}_2 (=6,3)$, goes like the original Fibonacci sequence, with returns to $D^{(2)}_1$ for every even $q$.
This is indeed true for any local symmetry range, yielding a third remarkable property:\\
(iii) For any maximal palindrome length $L_r$, the sequence of spacings $d_F \equiv (d^{(r)}_q)_{q\geqslant 1}$ is generated as
\begin{equation} 
d^{(k)}_F = \sigma_F^k(A), ~~ k=0,1,2,...,
\label{eq.fibspacing}
\end{equation}
just like the generations of $w_F$ in Eq.~(\ref{eq.fib}), but with the Fibonacci substitution rule $\sigma_F(A,B)$ now acting on the renormalized 'spacing alphabet'
\begin{equation}
\{A = D^{(r)}_3D^{(r)}_1,~ B = D^{(r)}_2D^{(r)}_1\},
\end{equation}
with additional starting letter $d_F[-1]=B$ (see Fig.~\ref{fig.fib-pal-spacing}(b)).
Note that the spacings $D^{(r)}_i$ are here treated simply as symbols being concatenated into a sequence.

Equivalently, the sequence $w_F$ can itself be renormalized by mapping its letters $B$ and (squares of) $A$ to the occurring spacings for any given $L_r$ as follows:
\begin{equation}
B \to D^{(r)}_1,~ A \to D^{(r)}_2, ~ A^2 \to D^{(r)}_3,
\label{eq.fib-renorm}
\end{equation}
with the $B$'s thus mapped to the spacing returned to at every second step.
Renormalized according to Eqs.~(\ref{eq.fib-renorm}), the original Fibonacci sequence is mapped exactly to the three-letter (or {\it trinary}) sequence $d_F$:
\begin{eqnarray}
w_F &=& A~B~A^2~B~A~B~A^2~B \cdots \nonumber \\ &\to& D_2D_1D_3D_1D_2D_1D_3D_1 \cdots = d_F.  
\label{eq.fib-renorm-seq}
\end{eqnarray}

From the above properties (i)--(iii), the total local symmetry (at different spatial scales) of the Fibonacci lattice can be seen to possess a {\it dynamical self-similarity}, which uniquely characterizes its long-range aperiodic order.

\subsection{Generalized Fibonacci lattices}

The Fibonacci inflation rule can be generalized \cite{Wang2000_PhysRevB.62.14020,Poddubny2010_Physica.E.42.1871} to powers of $A$ and $B$ as
\begin{equation} \sigma_{m,n}(A,B) = (A^mB^n,A) , \label{eq.fibgen} 
\end{equation}
where the word generations $w^{(k)}_{m,n}=\sigma^k_{m,n}(A)$ are equivalently given by the recursion scheme 
\begin{equation}w^{(k)}_{m,n} = [w^{(k-1)}_{m,n}]^m[w^{(k-2)}_{m,n}]^n, ~k \geqslant 2.
\end{equation}
Their lengths are, accordingly, $|w^{(k)}_{m,n}| = m|w^{(k-1)}_{m,n}|+n|w^{(k-2)}_{m,n}|$.
The generated infinite word $w_{m,n} = \lim_{k\to\infty}w^{(k)}_{m,n}$ then has limiting word length ratio \cite{Wang2000_PhysRevB.62.14020} 
\begin{equation}\varphi_{m,n} = \lim_{k\to\infty}\frac{|w^{(k+1)}|}{|w^{(k)}|} = \frac{m + \sqrt{m^2+4n}}{2}.
\end{equation}
For the standard Fibonacci word we thus have $\varphi_{1,1} = \varphi_g$.
For general letter powers $m,n$ the resulting lattice is aperiodic, while for $n=1$ it is also quasiperiodic \cite{Wang2000_PhysRevB.62.14020,Poddubny2010_Physica.E.42.1871}.
Having studied the local symmetries of the standard Fibonacci lattice, we now briefly address two further examples with quasiperiodic order.

\subsubsection{Silver Fibonacci lattice}

\begin{figure}[t]
\includegraphics[width=.88\columnwidth]{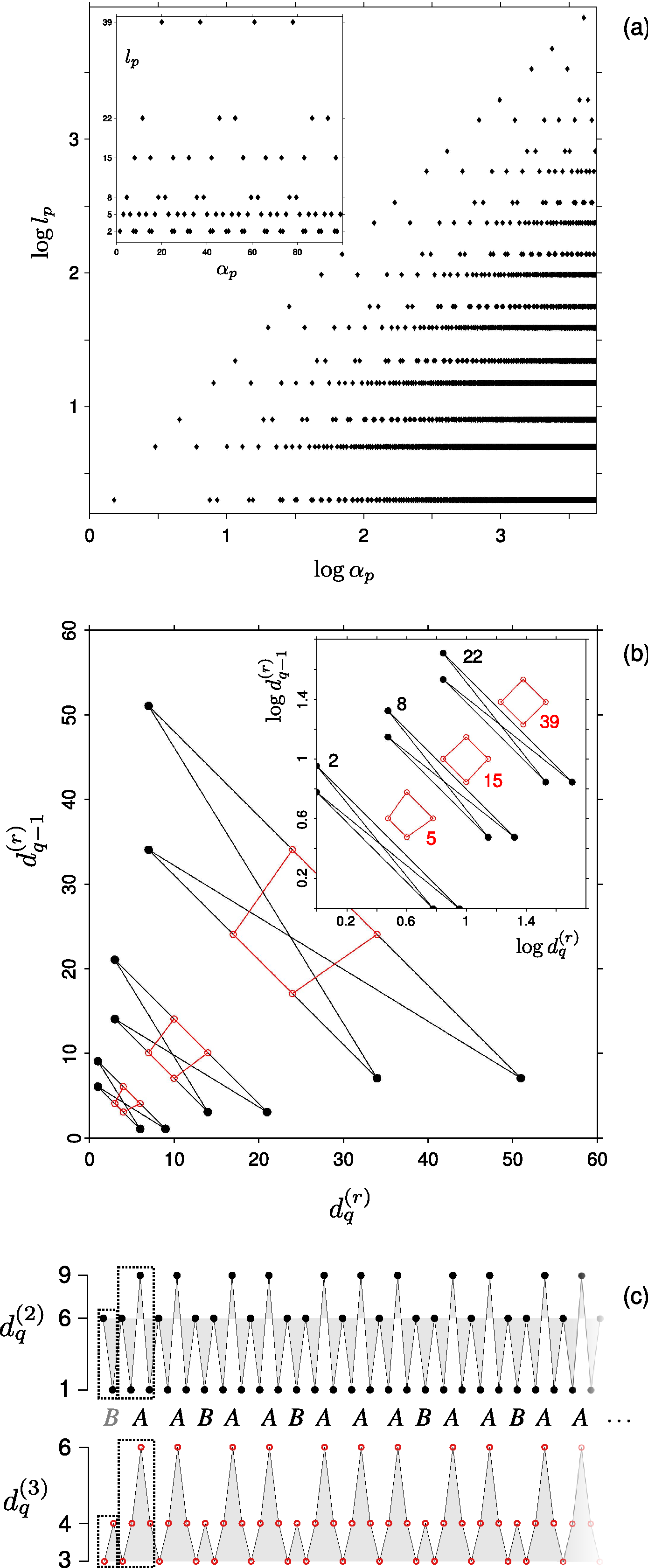}\vspace{.5cm}
\caption{\label{fig.fib21} (Color online) Silver Fibonacci sequence, $w_{2,1}$. (a) Local symmetry distribution (like in Fig.~\ref{fig.fib-pal-cl}). (b) Spacing return maps for different $L_r$, with full (empty) cirles for even (odd) $r$, repeated on logarithmic scale in the inset (like in Fig.~\ref{fig.fib-pal-spacing}(a)). (c) Spacing trajectories for even (top) and odd (bottom) $r$, represented by $d^{(2)}_q$ and $d^{(3)}_q$, respectively.}
\end{figure}

In Fig.~\ref{fig.fib21}(a) the local symmetry distribution ($\alpha_p,l_p$) is shown for the lattice $w_{2,1}$, with $\varphi_{2,1} = 1+\sqrt{2} = \varphi_s$ referred to as the {\it silver} mean.
Again, the first occurring palindromes $\pi^{(r)}_{q=1}$ for every given length $L_r = 1,2,5,8,15,22,...$ ($L_1 = 1$ not plotted) are prefixes; i.e., Eq.~(\ref{eq.prefix}) holds.
However, the $L_r$ now form two subsequences for odd and even $r$ having, as we will see, different structural properties.
For odd $r$, the $L_r$ are given by the lengths of the $r$-th order generated words with the last two letters discarded, 
\begin{equation}
L_{r=\rm odd} = |w^{(k=\frac{r+1}{2})}_{2,1}|-2,
\end{equation}
like in the standard Fibonacci case.
The ranges for even $r$ are asymptotically related to the above as 
\begin{equation}
\lim^{r=\rm even}_{r\to\infty}\frac{L_{r+1}}{L_{r}} = \varphi_{2,1}-1 = \sqrt{2}.
\end{equation}
Thus, for both odd and even sequences the ratio of consecutive prefix palindromes converges to the silver mean, $\lim_{r\to\infty} L_{r+2}/L_{r} = \varphi_{2,1}$, in analogy to Eq.~(\ref{eq.lambda}).

The local symmetry dynamics of $w_{2,1}$ is shown in Fig.~\ref{fig.fib21}(b), as represented by the return maps of the spacings $d^{(r)}_q$ for each symmetry range $L_r$.
For each $r$ we have three possible, non-repeated spacings $D^{(r)}_i$ ($i=1,2,3$), as in the standard Fibonacci case, now with 
\begin{equation}3D^{(r=\rm even)}_2 = 2D^{(r)}_3, ~~~ D^{(r=\rm odd)}_3 = 2D^{(r)}_1,
\end{equation}
which scale asymptotically as 
\begin{equation}\lim^{r=\rm even}_{r\to\infty}\frac{D^{(r)}_2}{D^{(r)}_1} = 2\varphi_{2,1}~, ~~~ \lim^{r=\rm odd}_{r\to\infty}\frac{D^{(r)}_2}{D^{(r)}_1} = \varphi_{2,1}-1.
\label{eq.fib21-spacing}
\end{equation}
For even (odd) $r$, there are consecutive returns to $D^{(r)}_1$ ($D^{(r)}_2$) in each map.
Again, the minimal spacings follow the numeric sequence of lengths of $w^{(k)}_{2,1}$, but now pairwise: 
\begin{equation}D^{(r=\rm odd)}_1 = D^{(r+1)}_1 = |w^{(k=\frac{r-1}{2})}_{2,1}|.
\end{equation}
The odd and even subsequences of symmetry ranges are further nested through the relations 
\begin{equation}
 D^{(r=\rm odd)}_3 = D^{(r-1)}_2.
\end{equation}
Also, the spacings in consecutive return maps obey, with $r$ odd, 
\begin{subequations}
\begin{align}D^{(r)}_1 + D^{(r)}_2 = D^{(r-1)}_1 + D^{(r-1)}_2, \\ D^{(r)}_2 + D^{(r)}_3 = D^{(r-1)}_1 + D^{(r-1)}_3,\end{align}  
\end{subequations}
so that the pairs $r$, $r-1$ have two lines with collinear points (see linear plot in Fig.~\ref{fig.fib21}(b)).

The presence of the $w_{2,1}$ aperiodic order in its detailed local symmetry dynamics is illustrated in Fig.~\ref{fig.fib21}(c) for $r=2$ and $r=3$.
Similarly to the golden Fibonacci case, the actual spacing trajectories $d^{(r)}_q$ for given $L_r$ coincide with the original aperiodic sequence $w_{2,1}$, when 'renormalized' and shifted by $\Delta q = 2$:
The corresponding sequence $d_{2,1}$ of spacing symbols $D^{(r)}_i$ is generated, in analogy to Eq.~(\ref{eq.fibspacing}), as $d^{(k)}_{2,1} = \sigma_{2,1}^k(A)$ ($k=0,1,2,...$) by the substitution rule $\sigma_{2,1}(A,B)$ on the alphabet
\begin{subequations}
\begin{align}
\{A &= D^{(r)}_1D^{(r)}_2D^{(r)}_3D^{(r)}_2, ~ B = D^{(r)}_1D^{(r)}_2\} \label{eq.fib-renorm-odd} \\ 
{\rm or}~~~\{A &= D^{(r)}_2D^{(r)}_1D^{(r)}_3D^{(r)}_1, ~ A = D^{(r)}_2D^{(r)}_1\} \label{eq.fib-renorm-even} 
\end{align}
\end{subequations}
for odd (\ref{eq.fib-renorm-odd}) or even (\ref{eq.fib-renorm-even}) $r$, respectively, again with additional starting letter $d_{2,1}[-1]=B$.
The grouping of spacings into letters is depicted by the dotted lines in Fig.~\ref{fig.fib21}(c).

Conversely, the letter sequence $w_{2,1}$ can be renormalized by mapping the single $B$'s and the squares and cubes of $A$ to the occurring spacings for given $L_r$,
\begin{subequations}
\begin{align}
B &\to D^{(r)}_1 ~~~~ (D^{(r)}_2),\\ 
A^2 &\to D^{(r)}_2 ~~~~ (D^{(r)}_1),\\ 
A^3 &\to D^{(r)}_3
\end{align} 
\end{subequations}
for even (odd) $r$.
The original sequence $w_{2,1}$ is then renormalized into the trinary sequence $d_{2,1}$ of the spacing trajectory,
\begin{eqnarray}
w_{2,1} &=& A^2~B~A^2~B~A^3~B~A^2~B \cdots \nonumber \\ &\to& D_2D_1D_2D_1D_3D_1D_2D_1 \cdots \nonumber \\ ( &\to& D_1D_2D_1D_2D_3D_2D_1D_2 \cdots) \equiv d^{e(o)}_{2,1} 
\label{eq.fib21-renorm-seq}
\end{eqnarray}
for even (odd) $r$; see Fig.~\ref{fig.fib21}(c).

This latter mapping of the original aperiodic sequence onto its local symmetry spacing dynamics, which was described already for the golden Fibonacci case (Eqs.~(\ref{eq.fib-renorm}) and (\ref{eq.fib-renorm-seq})), can be intuitively anticipated from the local symmetry distribution:
The pattern of symmetry spacings for certain maximal range is expected to follow, in some way, the pattern of isolated $B$'s among powers of $A$, which in turn is directly given in the original letter sequence--and indeed, this way is exactly provided by Eqs.~(\ref{eq.fib-renorm-seq}) and (\ref{eq.fib21-renorm-seq}).
Moreover, this renormalization scheme can be extended to larger $m$ in the class of quasiperiodic generalized Fibonacci lattices, as will be seen in the next subsection.

\subsubsection{Generalized quasiperiodic Fibonacci lattices}

\begin{figure}[t]
\includegraphics[width=.88\columnwidth]{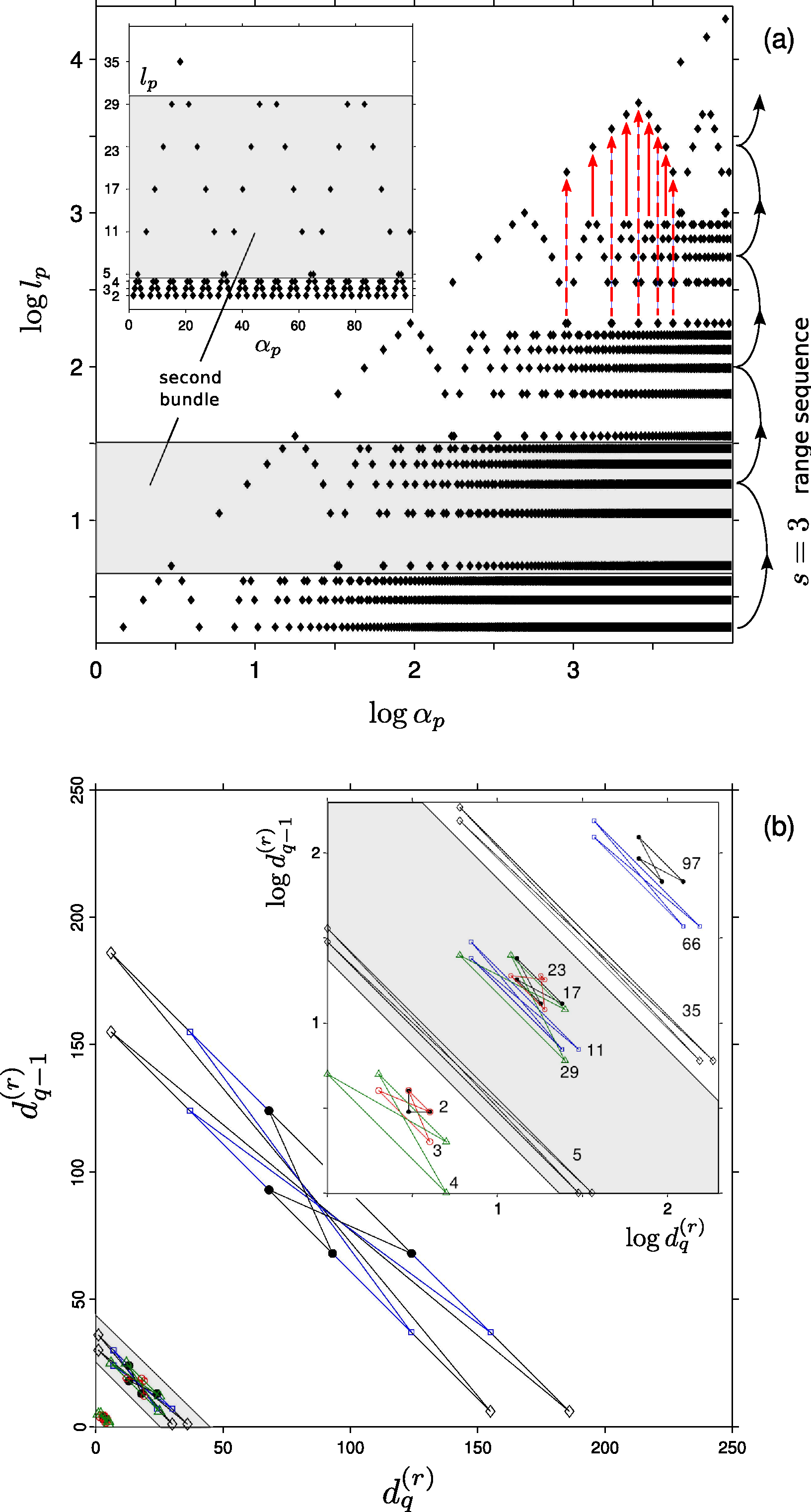}\vspace{.5cm}
\caption{\label{fig.fib51} (Color online) Generalized Fibonacci $w_{5,1}$ lattice. Local symmetry (a) distribution and (b) spacing return maps (like in Fig.~\ref{fig.fib21}(a) and (b), respectively). The curved arrows in (a) distinguish the $3^{\rm rd}$ subsequence of symmetry ranges (see text), and the shaded areas show (a) the $2^{\rm nd}$ scaling quintet bundle and (b) the corresponding spacing return maps. The straight vertical arrows in (a), starting from $\wedge$-like structures, show the jumps to larger symmetry ranges belonging to the next (solid lines) or next-to-next (dashed lines) quintet bundle (see text).}
\end{figure}

As we have seen above for the quasiperiodic case $n=1$, the $m=1$ (golden) and $m=2$ (silver) sequences feature one and two characteristic asymptotic scaling(s) for the spacing of local symmetries, respectively.
This scheme continues for larger $m$, so that '$m$-tets' of asymptotic spacing scalings arise.
We will use the index $s = 1,2, ..., m$ to count the members of an $m$-tet:
The $s$-th spacing scaling in an $m$-tet corresponds to the subsequence composed of every $m$-th occurring maximal local symmetry range, starting with $L_{r=s-1}$ (e.g., for $s=3$ we get the third such subsequence, $L_{2},L_{2+m},L_{2+2m,...}$, marked in Fig.~\ref{fig.fib51}(a)) \cite{bundles}.
Equal-order members of these $m$ range-subsequences form '$m$-tet bundles' (e.g., the bundle of second members of the $m=5$ subsequences is depicted in Fig.~\ref{fig.fib51}(a)).
For any $m$, the first occurring palindromes for given ranges $L_r$ are prefixes of $w_{m,1}$.
Also, although in general the $m$-tet subsequences have different palindrome {\it spacing} scalings, their palindrome {\it lengths} share the same scaling, given by the inverse slope of $w_{m,1}$ (like the cases $m=1,2$ seen above): 
\begin{equation}\lim_{r\to\infty} \frac{L_{r+m}}{L_{r}} = \varphi_{m,1}.
\end{equation}
Further, the last length in each $m$-tet bundle follows the length sequence of $w_{m,1}$ in the same way as seen so far for $m=1$ and $2$ (with $k \geqslant 2$ for $m=1$),
\begin{equation}L_{r=km-1} = |w^{(k)}_{m,1}| - 2 ~~~(k = 1,2, ...)~.
\end{equation}

As an example, we briefly consider the case $(m=5, n=1)$, i.e. the sequence $w_{5,1}$.
In Fig.~\ref{fig.fib51}(a), its local symmetry distribution is plotted, showing a grouping of the ranges $L_r$ into quintets (horizontal bundles at increasing scale, with five ranges each).
It is evident how, due to the recurrent $A^m$-factors, axes of maximal palindromes located at the left and right ends of a larger palindrome are shifted inwards for increasing $L_r$, thereby forming characteristic $\wedge$-like structures, along the sequence (i.e., the $\alpha_p$-axis) and at increasing scales (i.e., the $l_p$-axis).
At the meeting point $\alpha_p$ of the legs of each such $\wedge$, a larger local symmetry occurs, with corresponding range $L_r$ belonging to the next or next-to-next $m$-tet bundle (as indicated by vertical arrows in Fig.~\ref{fig.fib51}(a)); 
see also Fig.~\ref{fig.fib101-fib110-pal-cl}(a), showing the bundle structure of the $w_{10,1}$ sequence.

The spacing dynamics for the different symmetry ranges become increasingly complex for higher $m$, as can be seen from the return maps for $m=5$ in Fig.~\ref{fig.fib51}(b).
Just as the odd-even maps in the $m=2$ case, the maps corresponding to $L_r$'s of the same $m$-tet bundle in Fig.~\ref{fig.fib51}(a) are collinear, giving evidence to a nested order of symmetry dynamics along the lattice.
Out of the $m$ (different) spacing scalings of the $m$-tet, the asymptotic scaling for the {\it leading} subsequence (the sequence of smallest symmetry ranges in subsequent $m$-tet bundles) is directly related to the slope of $w_{m,1}$ as (cf. cases $m=1,2$ above, Eqs.~(\ref{eq.fib-spacing}), (\ref{eq.fib21-spacing}))
\begin{equation}
\lim^{r=km-1}_{k\to\infty}\frac{D^{(r)}_2}{D^{(r)}_1} = m\varphi_{m,1} = 5\varphi_{5,1} = 5 \frac{(5+\sqrt{29})}{2}.\end{equation}                                                                                                                          

Finally, as mentioned above, suitable renormalization of the original sequence $w_{m,1}$, by mapping the single $B$'s and powers of $A$ to the spacings $D^{(r)}_i$, again yields the spacing trajectory.
In particular, for all $L_r$ contained in any $s$-th range-subsequence ($s=1,2,...m$), there exists a common mapping 
\begin{equation}
 B \to D^{(r)}_i, ~ A^m \to D^{(r)}_j, ~ A^{m+1} \to D^{(r)}_k,
\end{equation}
so that 
\begin{equation}
 w_{m,1} \to d^{(s)}_{m,1} \equiv (d^{(r)}_q)_{q\geqslant 1}
\end{equation}
with the different $i,j,k \in \{1,2,3\}$ depending on the chosen $s$ (cf. cases $m=1,2$ above, Eqs.~(\ref{eq.fib-renorm-seq}), (\ref{eq.fib21-renorm-seq})).

\subsubsection{Copper Fibonacci lattice}

\begin{figure}[t]
\includegraphics[width=.88\columnwidth]{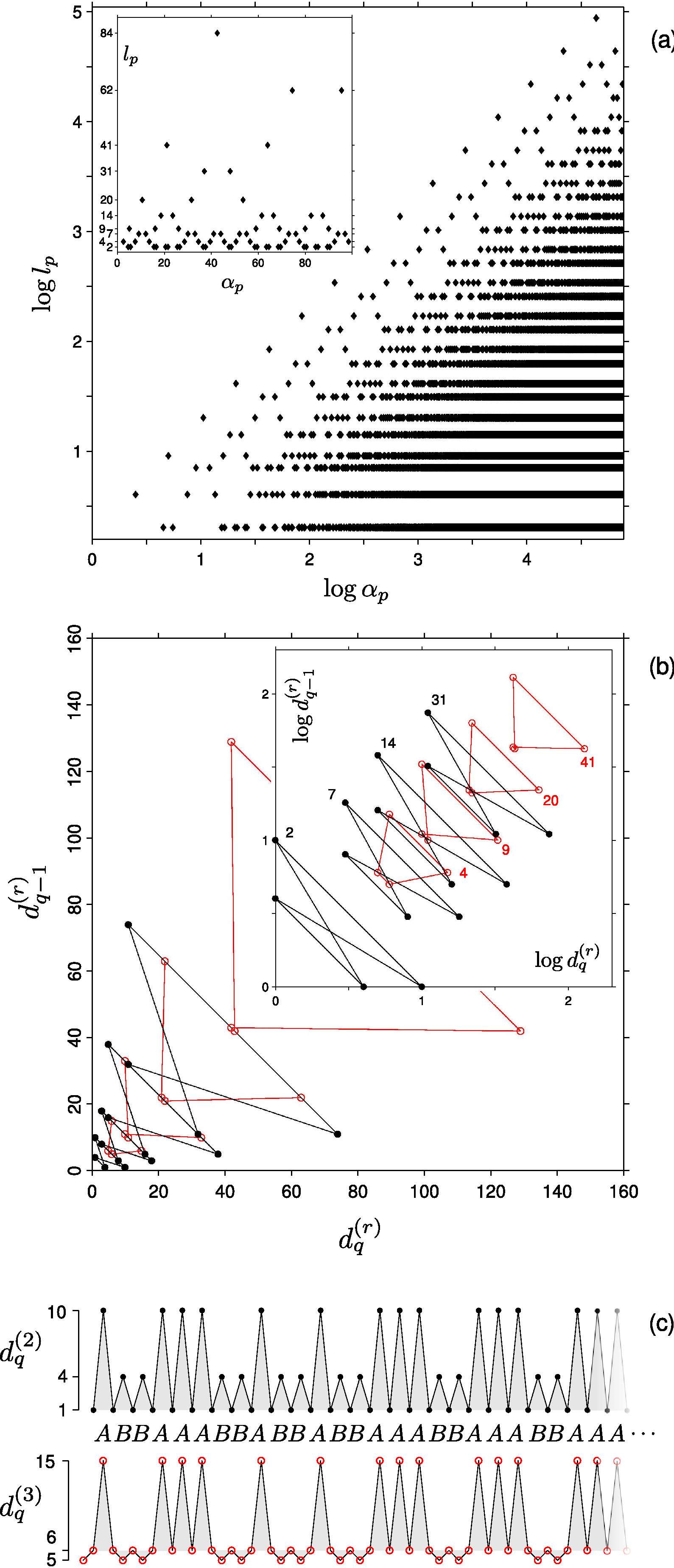}\vspace{.5cm}
\caption{\label{fig.fib12} (Color online) Copper Fibonacci lattice, $w_{1,2}$. Local symmetry (a) distribution, (b) spacing return maps and (c) spacing trajectories for even and odd $r$ (like in Fig.~\ref{fig.fib21}(a), (b) and (c), respectively).}
\end{figure}

As an example of a non-quasiperiodic generalized Fibonacci sequence (i.e., $n>1$ in Eq.~(\ref{eq.fibgen})) we consider the case $m=1, n=2$ in Eq.~(\ref{eq.fibgen}).
The lengths of the word generations $w^{(k)}_{1,2}$ are recursively given by $|w^{(k)}_{1,2}| = |w^{(k-1)}_{1,2}|+2|w^{(k-2)}_{1,2}|$, and have limiting ratio $\varphi_{1,2} = \varphi_c = 2$, known as the {\it copper} mean.
The local symmetry distribution of $w_{1,2}$ is shown in Fig.~\ref{fig.fib12}(a).
Like in the $w_{2,1}$ case, the sequence of occurring maximal palindrome lengths, $(L_r)_{r \geqslant 1} = 1, 2, 4, 7, 9, 14, 20, ...$ ($L_1 = 1$ not plotted), is split into two subsequences for odd and even $r$.
Here, though, for the subsequence of even $r$ the first occurring maximal palindromes (with axes $\alpha^{(r)}_{q=1}$) are {\it not} prefixes of $w_{1,2}$, that is, no complete local symmetries of those ranges are encountered.

Further, the two $L_r$ subsequences are related to the length sequence $|w^{(k)}_{1,2}|$ in a more involved way, through their {\it difference} $\Delta L_r \equiv L_{r+1} - L_r$, $r \geqslant 2$, as follows.
For odd $r$, $\Delta L_r$ coincides with the sequence of lattice generation lengths, while for even $r$ it oscillates around it by alternately  $\pm 1$:\\
\begin{subequations}
\begin{align}
\Delta L_{r=\rm odd} &= |w^{(k=\frac{r-1}{2})}_{1,2}| = \nonumber \\ 
&= |w^{(0)}_{1,2}|, |w^{(1)}_{1,2}|, |w^{(2)}_{1,2}|, ... \nonumber \\ &= 1,3,5,11,21,43,85,..., \\ \newline
\Delta L_{r=\rm even} &= |w^{(k=\frac{r-2}{2})}_{1,2}| + (-1)^{\frac{r{\rm mod}4}{2}+1} \nonumber \\ 
&= |w^{(0)}_{1,2}|+1, |w^{(1)}_{1,2}|-1, |w^{(2)}_{1,2}|+1, ... \nonumber \\  &= 2,2,6,10,22,42,86,...~.
\end{align}
\end{subequations}
As we see, the non-quasiperiodic $w_{1,2}$ sequence features more indirect connection of its palindrome distribution to the structure of the original letter sequence than the previous cases (i.e., the quasiperiodic $w_{m,1}$ sequences).
Its asymptotic palindrome length scaling, on the other hand, is the same for the odd and even $r$ subsequences, 
\begin{equation}
\lim_{r\to\infty} \frac{L_{r+2}}{L_{r}} = \varphi_C = 2 
\label{eq.fib12-length-scaling}
\end{equation}
with asymptotic even-to-odd-$r$ ratio 
\begin{equation}
 \lim^{r={\rm odd}}_{r\to\infty} \frac{L_{r+1}}{L_{r}} = \frac{3}{2}.
\label{eq.fib12-length-ratio}
\end{equation}

The spacing return maps for different local symmetry ranges in $w_{1,2}$ are shown in Fig.~\ref{fig.fib12}(b).
There are again $3$ possible spacings in each map, and now any map point for $k \geqslant 3$ lies on an antidiagonal line with $8$ collinear points, $4$ for each subsequence (odd and even $r$) of lengths $L_r$.
This connects the consecutive maps of each subsequence, but also fixes the subsequences to each other, as seen in the linear plot of Fig.~\ref{fig.fib12}(b).
The asymptotic scalings of the odd and even $r$ spacings with respect to the $D^{(r)}_1$ in the maps is 
\begin{equation}
\lim^{r=\rm odd, even}_{r\to\infty}\frac{D^{(r)}_2}{D^{(r)}_1} = 1,~3
\end{equation}
for the middle spacings, and 
\begin{equation}
\lim^{r=\rm odd, even}_{r\to\infty}\frac{D^{(r)}_3}{D^{(r)}_1} = 3,~7
\end{equation}
for the maximal spacings.

Also here the renormalized spacing trajectories for given $L_r$ coincide with the original aperiodic sequence $w_{1,2}$:
The sequence of spacings is given iteratively by $d^{(k)}_{1,2} = \sigma_{1,2}^k(A)$ on the alphabet 
\begin{subequations}
 \begin{align}
\{A = D^{(r)}_1D^{(r)}_3, B = D^{(r)}_1D^{(r)}_2\} \\ {\rm or}~~~ \{A = D^{(r)}_3D^{(r)}_2, B = D^{(r)}_1D^{(r)}_2\}
 \end{align}
\end{subequations}
for odd or even $r$, respectively, with additional starting letter $d_{1,2}[-1]=B$.
This is shown for the $r=2$ and $r=3$ trajectories in Fig.~\ref{fig.fib12}(c).

\begin{figure}[t]
\includegraphics[width=.85\columnwidth]{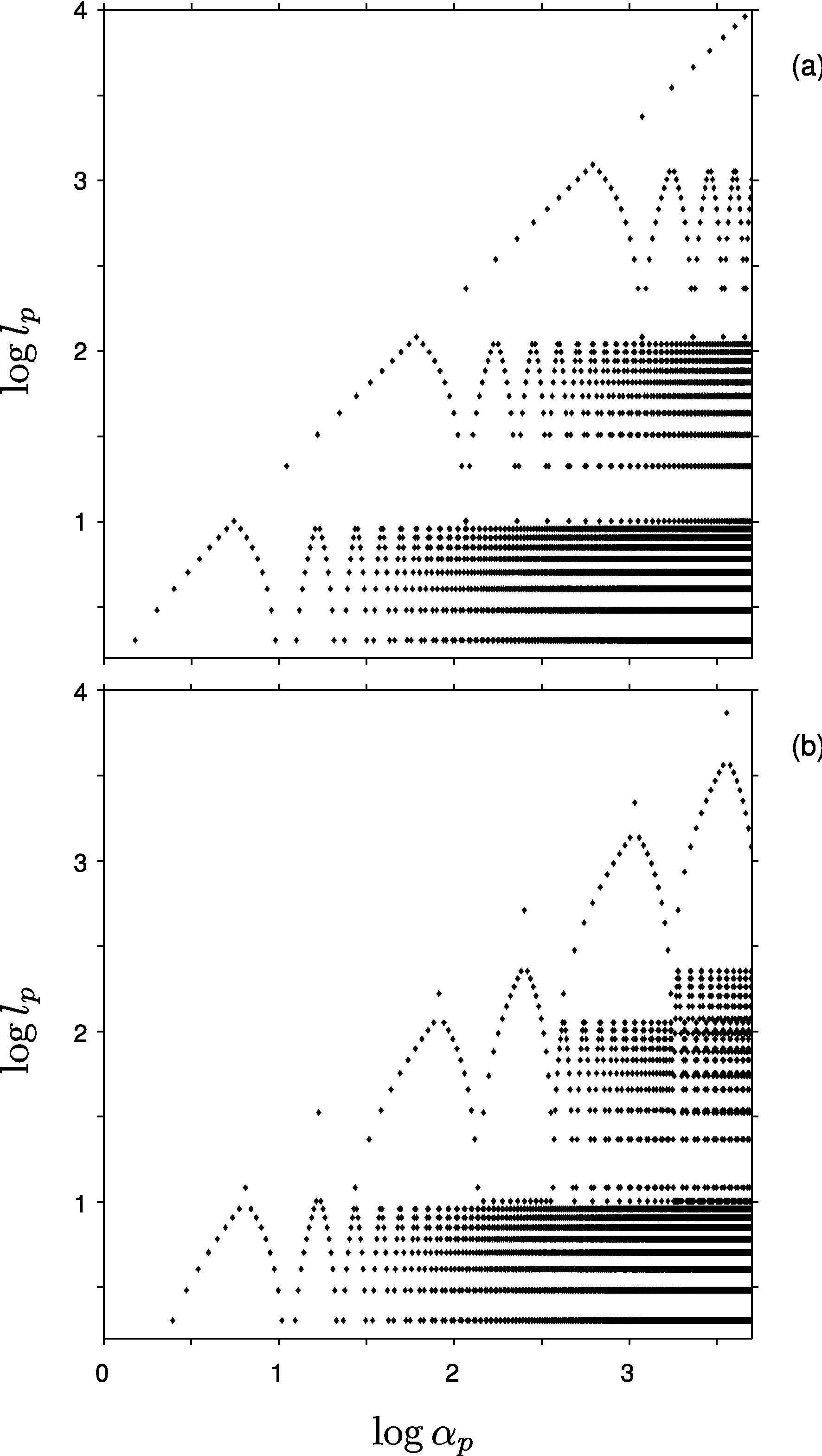}\vspace{.5cm}
\caption{\label{fig.fib101-fib110-pal-cl} Local symmetry distributions of the generalized Fibonacci lattices (a) $w_{10,1}$, with separate $m$-tet (here $m=10$) bundles forming along the $\alpha_p$-axis (in analogy to Fig.~\ref{fig.fib51}(a)), and (b) $w_{1,10}$, with intertwined $n$-tet (here $n=10$) bundles (see text).}
\end{figure}

To illustrate the characteristic differences between the local symmetry properties of the quasiperiodic ($n=1$) and non-quasiperiodic ($n>1$) generalized Fibonacci lattices for larger $m$ and $n$, Fig.~\ref{fig.fib101-fib110-pal-cl} shows the maximal palindrome distributions for the sequences $w_{10,1}$ and $w_{1,10}$.
It is clearly seen how $m$- and $n$-tets of maximal symmetry range subsequences occur (in Fig.~\ref{fig.fib101-fib110-pal-cl}(a) and (b), respectively), which form bundles containing consecutive (in $ \alpha_p$-direction) $\wedge$-like structures of approaching symmetry axes for increasing $L_r$, like in Fig.~\ref{fig.fib51}(a).
At the meeting points of the legs of the $\wedge$'s larger symmetry ranges occur, belonging to a subsequent (in $l_p$-direction) $m$- or $n$-tet bundle.
In the quasiperiodic case, Fig.~\ref{fig.fib101-fib110-pal-cl}(a), the first maximal palindrome of each occurring length (i.e., all points on the left leg of each first-occurring $\wedge$) is a prefix of $w_{m,1}$, and the $m$-tet bundles are separated.
On the contrary, in the non-quasiperiodic case, Fig.~\ref{fig.fib101-fib110-pal-cl}(b), prefix palindromes occur only at the meeting points of first-occurring $\wedge$'s, and the $n$-tet bundles are, in general, intertwined (i.e., bundles overlap with successors containing different $L_r$'s).

\subsection{Period doubling lattice}

\begin{figure}[t]
\includegraphics[width=.88\columnwidth]{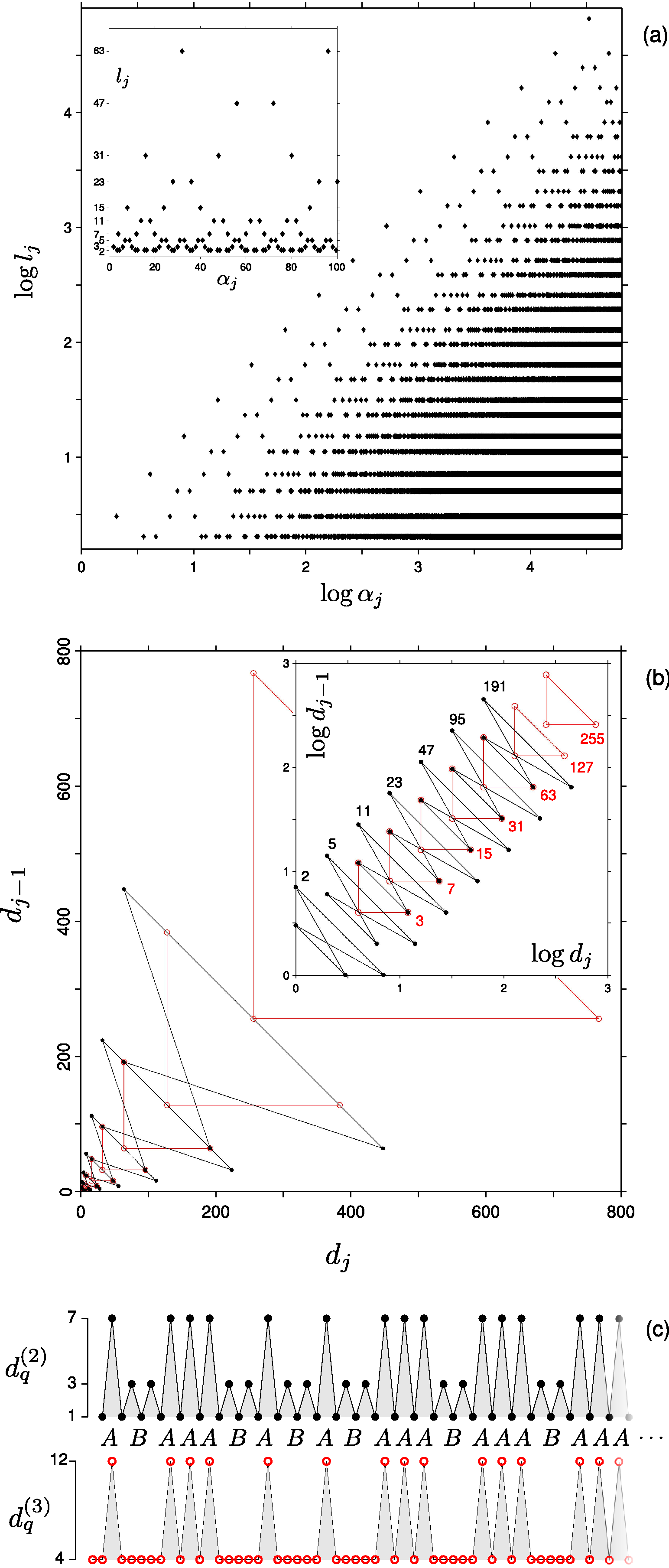}\vspace{.5cm}
\caption{\label{fig.pd} (Color online) Period doubling lattice, $w_P$. Local symmetry (a) distribution, (b) spacing return maps and (c) spacing trajectories for even and odd $r$ (like in Fig.~\ref{fig.fib21}(a), (b) and (c), respectively).}
\end{figure}

Let us now inspect the local symmetries of the period-doubling (PD) lattice, whose extensively studied structural and spectral properties  \cite{Belissard1991_Comm.Math.Phys.135.379,Hof1995_Comm.Math.Physics.174.149,Damanik1998_Commun.Math.Phys.196.477,Damanik2000_Disc.Appl.Math.100.115,Wang2000_PhysRevB.62.14020,Allouche2003_Th.Comp.Sc.292.9,Baake2010_J.Phys.Conf.Ser.226.012023} manifest its self-similar nature, although it is not categorized as quasicrystalline \cite{Macia2006_RepProgPhys.69.397, Poddubny2010_Physica.E.42.1871}.
The PD letter sequence $w_P = \lim_{k\to\infty} w^{(k)}_P$, with $w^{(k)}_P = \sigma_P^k(A)$, is generated by the inflation rule \begin{equation}
\sigma_P^k(A,B)=(AB,AA),
\end{equation}
that is, like a standard Fibonacci rule but with a squared image of $B$, and the length of the $k$-th generation word is now $|w^{(k)}_P| = 2^k$.
Its recursion scheme is $w_P^{(k)}=w_P^{(k-1)}[w_P^{(k-2)}]^2$, which coincides with that of the copper Fibonacci sequence $w_{1,2}$.
In fact, $w_P$ is obtained from $w_{1,2}$ simply by substituting all squares $BB$ by single $B$'s.
We can thus expect these lattices to have certain structural similarities.

In Fig.~\ref{fig.pd}(a) the local symmetry distribution of $w_P$ is shown, which, indeed, has the same characteristics as the distribution for $w_{1,2}$ described above.
Again, the sequence of occurring lengths, $L_r$, consists of two subsequences for odd and even $r$ (i.e., doublets of local symmetry scaling), the former of which contains all palindromic prefixes.
The lengths of the palindromes in the odd and even subsequences relate to the $w_P$ word generation lengths as 
\begin{subequations}
\label{eq.pd-length}
\begin{align}
L_{r=\rm odd} &= |w^{(k=\frac{r+1}{2})}_P| - 1 = 2^k - 1\\ 
L_{r=\rm even} &= \frac{3}{2}|w^{(k=\frac{r}{2})}_P| - 1 = \frac{3}{2}\cdot 2^k - 1,
\end{align}
\end{subequations}where $k=1,2,...$..
From these relations it is obvious that the maximal palindrome lengths in each (odd and even $r$) subsequence scale asymptotically as the original PD sequence, $\lim_{r\to\infty} L_{r+2}/L_r = 2$, with $\lim^{r=\rm odd}_{r\to\infty} L_{r+1}/L_{r} = 3/2$, just like for the copper Fibonacci sequence (Eqs.~(\ref{eq.fib12-length-scaling}) and (\ref{eq.fib12-length-ratio}), respectively).

Moreover, it is clear that each palindrome prefix, having odd length $L_r$, contains the previous prefix palindrome of length $L_{r-2}$ (for odd $r \geqslant 3$) in its left half, i.e., up to (but not including) its central letter.
Specifically, it is composed as 
\begin{equation}\pi' = \pi x \pi; ~~|\pi'|=L_r, ~ |\pi|=L_{r-2}, ~ x \in \{A,B\}.
\end{equation}
Simultaneously, though, its prefix $\pi x$ coincides with a word generation $w^{(k)}_P$, which ends with the letter $A$ ($B$) for even (odd) $k$.
This means that the PD sequence can be constructed solely from its local symmetries, by successively creating larger prefix palindromes from the repetition of the previous prefix palindrome with an additional central letter $x_k$, which follows a period $1$ oscillating sequence, $(x_k)_{k\geqslant 0}=A,B,A,B,...$:
Beginning with $\pi_0 = x_0 \equiv w^{(0)}_P = A$, we have $\pi_1 = \pi_0 x_1 \pi_0 = ABA$, $\pi_2 = \pi_1 x_2 \pi_1 = ABAAABA$, ..., and, in general,
\begin{equation} \pi_k = \pi_{k-1} x_k \pi_{k-1} ~~~ (k\geqslant 1). 
\end{equation}
In this way, we obtain the relation 
\begin{equation} w^{(k)}_P = \pi_{k-1}x_k,
\end{equation}
as a recursive palindromic construction of the PD sequence via the 'directive' periodic sequence $(x_k)$.

The palindrome complexity function of the PD sequence, i.e. the number of different contained palindromes of length $l$, has been explicitly computed in Ref.~\cite{Damanik2000_Disc.Appl.Math.100.115}, and is given by 
\begin{subequations}
 \begin{align}
p_{w_P}(l={\rm odd}\geqslant 5) &= p_{w_P}(2l-1) = p_{w_P}(2l+1), \\ p_{w_P}(l={\rm even}\geqslant 4) &= 0, 
 \end{align}
\end{subequations}
with starting palindromes $\mathcal{P}_1(w_P)=\{A,B\}$, $\mathcal{P}_2(w_P)=\{AA\}$ (this being the only palindrome with even length), $\mathcal{P}_3(w_P)=\{AAA,ABA,BAB\}$, $\mathcal{P}_5(w_P)=\{AABAA,ABABA,BAAAB,BABAB\}$, and $\mathcal{P}_7(w_P)=\{AAABAAA,ABAAABA,ABABABA\}$.
Nevertheless, we note that, for any odd $l\geqslant 5$, the palindromes of length $2l-1$ are not {\it maximal} (see the inset of Fig.~\ref{fig.pd}(a)), that is, 
\begin{equation}\mathcal{M}^{(l={\rm odd}\geqslant 5)}_{2l-1}(w_P)=\varnothing.
\end{equation}
Further, even for lengths $l\in\{L_r\}$, not {\it all} palindromes are maximal; e.g., $\mathcal{M}_3(w_P)=\{ABA\}$, $\mathcal{M}_5(w_P)=\{ABABA\}$, and $\mathcal{M}_7(w_P)=\{ABAAABA\}$ (since $BB \notin \mathcal{P}(w_P)$ and $p_{w_P}(4)=0$).

The return maps of the local symmetry dynamics of the PD lattice are shown in Fig.~\ref{fig.pd}(b).
Their structure very much resembles, qualitatively, the maps for the copper $w_{1,2}$ sequence:
The subsequences of maps for lengths $L_r$ with odd and even $r$ are intertwined in exactly the same way by antidiagonally collinear map points.
Quantitatively, however, the PD local symmetry spacings evolve more regularly along the lattice than for the copper Fibonacci, with characteristics which are invariant (as opposed to asymptotic) in increasing $r$, as follows:
For all odd $r$ (red circles in Fig.~\ref{fig.pd}(b)), there are now two occurring spacings $D^{(r)}_1,D^{(r)}_2$ (and not {\it asymptotically} two, as in $w_{1,2}$) with repeated $D^{(r)}_1$, and three non-repeated spacings for even $r$.
In each subsequence, all spacings are {\it doubled} in the subsequent symmetry range, 
\begin{equation} D^{(r+2)}_i = 2D^{(r)}_i. 
\end{equation}
Further, the two map subsequences are fixed to each other by the relations 
\begin{equation}D^{(r = \rm odd)}_i = D^{(r+3)}_i,~i=1,2,
\end{equation} 
so that each odd map has two common points with the next-to-next even one.
The scaling of local symmetry spacings along the range sequences is also fixed as $D^{(r)}_2/D^{(r)}_1 = 3$ for all $r$ and $D^{(r)}_3/D^{(r)}_1 = 7$ for even $r$.

Finally, the spacing trajectories for even and and odd $r$ are shown in Fig.~\ref{fig.pd}(c).
Again the (renormalized) trajectories are generated by the original inflation rule $\sigma_P$.
For even $r$, we have the alphabet mapping 
\begin{equation}\{A = D^{(r)}_1D^{(r)}_3, B = D^{(r)}_1D^{(r)}_2D^{(r)}_1D^{(r)}_2 \};
\end{equation}
for odd $r$ it is the same but with $D^{(r)}_1 = D^{(r)}_2$, and with additional starting spacing $d_P[0] = d^{(r)}_0=D^{(r)}_1$.

\subsection{Thue-Morse Lattice}

\begin{figure}[t]
\includegraphics[width=0.88\columnwidth]{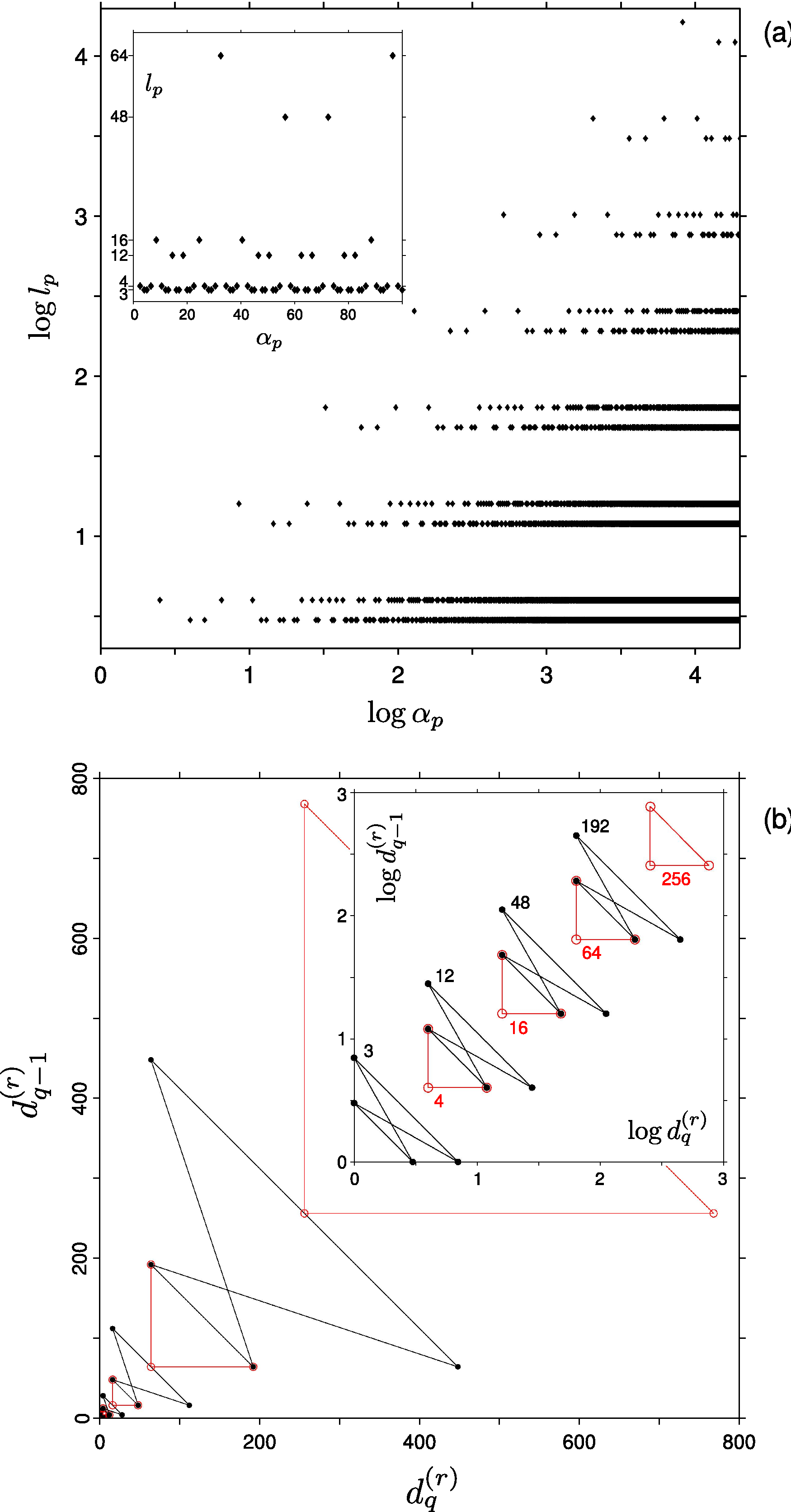}\vspace{.5cm}
\caption{\label{fig.tm} (Color online) Thue-Morse lattice, $w_T$. Local symmetry (a) distribution and (b) spacing return maps (like in Fig.~\ref{fig.pd}(a) and (b), respectively). Spacing trajectories coincide with the ones in Fig.~\ref{fig.pd}(c).}
\end{figure}

We now turn to another 1D lattice case with doublet local symmetry sequences, corresponding to the well-known (standard) Thue-Morse (TM) sequence, $w_T$, which shares the non-quasicrystalline spectral characterization of the PD lattice \cite{Poddubny2010_Physica.E.42.1871,Macia2006_RepProgPhys.69.397}.
Identifying $\{A,B\}$ with \{even,odd\}, $w_T$ can be defined as the parity sequence of the sum of digits of base-$2$ represented integers, but is otherwise ubiquitously present in various fields of mathematics and physics \cite{Allouche1999_Disc.Math.Th.Comp.Sc.1}.

As a symbolic sequence, $w_T$ is generated by the inflation rule 
\begin{equation}\sigma_T(A,B)=(AB,BA),
\end{equation}
and can be generalized in a similar manner as for the Fibonacci substitutions studied above \cite{Wang2000_PhysRevB.62.14020}, see Eq.~(\ref{eq.fibgen}).
The corresponding recursive concatenation scheme now has a two-component form \cite{Wang2000_PhysRevB.62.14020}
\begin{equation}w^{(k)}_T = w^{(k-1)}_T v^{(k-1)}_T, ~v^{(k)}_T = v^{(k-1)}_T w^{(k-1)}_T
\end{equation} 
for $k \geqslant 1$, starting with $w^{(0)}_T = A$, $v^{(0)}_T = B$, and the length sequence coincides with that of $w_P$:
$|w^{(k)}_T| = 2^k$.
The combinatorial palindromic properties of the TM sequence have, among others, been studied in Refs.\cite{Masse2008_Proc.GASCOM.2008.53,Masse2008_PUMA.19.39,Masse2008_IFIP.273.101}, where its palindrome complexity is also given, and connected to its Hamiltonian spectrum in Ref.\cite{Damanik2001_Ann.H.Poincare.2.927}.

Like for the PD sequence, we see that the standard TM substitution is very similar to the copper rule $\sigma_{1,2}$, where now $B$ is sent to the 'conjugate' image $BA$ of $A$, instead of the square $AA$.
On the other hand, $\sigma_T$ leads to the occurrence of the conjugate square $BB$ in $w_T$.
As a consequence, the local symmetries in $w_T$ will have ranges different from $w_P$, but partly feature similar structural characteristics, with depleted number of symmetry ranges due to the absence of the cube $AAA$.

In Fig.~\ref{fig.tm}(a) the local symmetry distribution of $w_T$ is shown.
Indeed, we see again a doublet structure of two symmetry range subsequences $(L_r)$, but now every second doublet bundle is missing compared to $w_P$ or $w_{1,2}$.
This is also reflected in the length sequences themselves, which are 'squared' with respect to the $w_P$ case (cf. Eq.~(\ref{eq.pd-length})):
\begin{subequations}
 \begin{align}
L_{r=\rm odd} &= |w^{(k=\frac{r-1}{2})}_T|^2 = 2^{2k}, \\ L_{r=\rm even} &= \frac{3}{2^2} |w^{(k=\frac{r}{2})}_T|^2 = \frac{3}{2^2}\cdot 2^{2k},
 \end{align}
\end{subequations}
with $k=1,2,...$.
There is now no 'residue' letter(s) to reach (a multiple of) the $w^{(k)}_T$ generation lengths (like the $-1$ for $w_P$ in Eq.~(\ref{eq.pd-length})), so that both subsequences have the same, constant scaling throughout the lattice, $L_{r+2}/L_r = 2^2$.
Also, in contrast to the PD case, here all $L_r$ are even except for the second one $L_2 = 3$ (and, of course, $L_1 = 1$).

The depletion of local symmetry ranges in the TM lattice is also evident in the spacing return maps, shown in Fig.~\ref{fig.tm}(b):
Every second return map in each subsequence is now missing with respect to the PD case, so that no 'overlap' between maps occurs, and the spacings are accordingly multiplied by four at subsequent symmetry ranges, $D^{(r+2)}_i = 2^2 D^{(r)}_i$.
For the remaining symmetry ranges, the (fixed) spacing scaling, $D^{(r)}_2/D^{(r)}_1 = 3$, $D^{(r=\rm even)}_3/D^{(r)}_1 = 7$, is identical to the PD lattice.

Remarkably enough, identical are also their spacing trajectories.
That is, despite the different inflation rules $\sigma_T$ and $\sigma_P$, and although the $L_r$ in $w_T$ are different (one letter larger) from the corresponding ones $w_P$, the spacings in the TM case are exactly the same as for the PD case, and follow the trajectories shown in Fig.~\ref{fig.pd}(c) for odd and even $r$.
This suggest a subtle (renormalization) relation between the TM and PD lattice, so that the former is also structurally conform to its maximal palindrome spacing trajectories.
Indeed, the TM lattice $ABBABAABBAABABBA\cdots$ can be transformed to the PD lattice \cite{Baake2010_J.Phys.Conf.Ser.226.012023} by the ($2$-to-$1$, surjective) map
\begin{equation} AA,BB \to A;~~ AB,BA \to B. 
\end{equation}
This more or less 'hidden' relation between the two maps is here directly (i.e., prior to any renormalization) reflected in the coincidence of their local symmetry dynamics.

\subsection{Cantor lattice}

\begin{figure}[t]
\includegraphics[width=0.92\columnwidth]{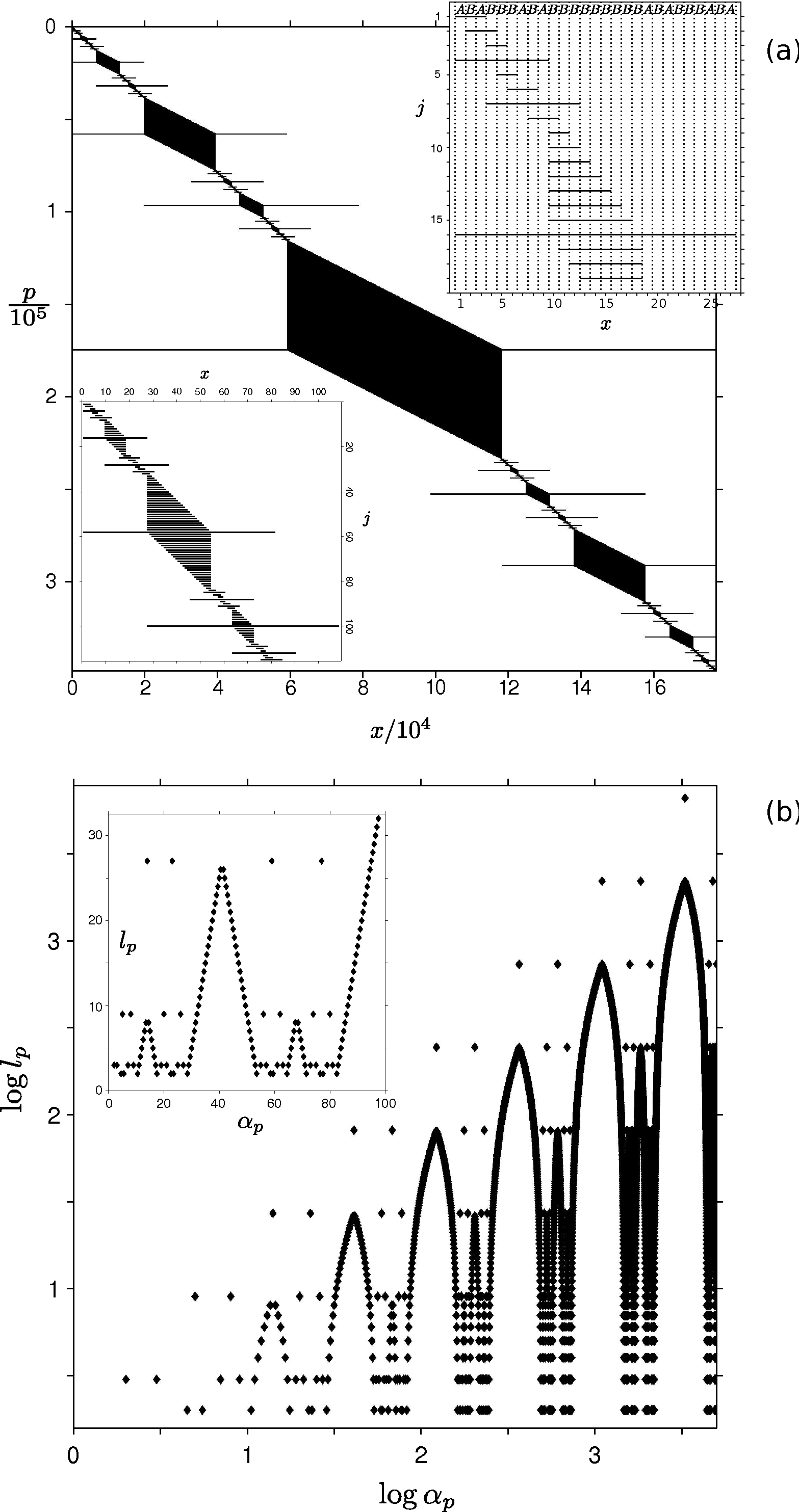}\vspace{.5cm}
\caption{\label{fig.cantor} Cantor lattice, $w_C$. (a) Maximal palindromes in order of occurrence in the letter sequence, the beginning of which is shown in the upper inset. The magnification in the lower inset demonstrates the self-similarity of the lattice. (b) Local symmetry distribution (like in Fig.~\ref{fig.fib-pal-cl}).}
\end{figure}

Having analyzed the local symmetry distribution and dynamics of some typical binary lattices with aperiodic order, we finally examine an extreme case with inherent palindromic self-similarity, the standard Cantor lattice.
In general, a 1D ($m$-fold) Cantor construct is obtained by dividing a finite segment into ($m$ equal) parts, 'deleting' a chosen number ($n<m$) of them which are not consecutive, and repeating this procedure recursively on each remaining segment ad infinitum (thus generating an '$m$-by-$n$' Cantor fractal).
Mapping the resulting full and empty intervals onto a binary alphabet ($A$'s and $B$'s, respectively), an infinite fractal letter sequence is obtained.
The sequence can be obtained equivalently by a substitution rule on the letters corresponding to the above partitioning scheme.
We here consider the homogeneous $3$-by-$1$ (or 'ternary') Cantor lattice \cite{Sengupta2004_Physica.B.344.307,Esaki2009_PhysRevE.79.056226}, which is obtained by recursively applying the inflation rule \begin{equation}\sigma_C(A,B)=(ABA,BBB)
\end{equation}
with starting letter $A$, yielding the sequence $w_C$.

Obviously, $\sigma_C$ inherently generates palindromes on all scales with ever increasing lengths, distributed self-similarly on a globally symmetric lattice.
The hierarchical local symmetry structure is clearly evident in Fig.~\ref{fig.cantor}(a), where the maximal palindromes of $w_C$ are plotted as horizontal segments, ordered in axis position.
In contrast to the (inhomogeneous) sequences seen so far, $w_C$ has a well defined global center, around which the primary cluster of $B$'s is located, covering the central third of the total lattice.
Secondary $B$-clusters are created around the centers of the outer thirds, and then on the outer thirds of these, etc.
Therefore, there will be maximal palindromes of increasing length with axes approaching these hierarchical symmetry centers, at which their length jumps by a factor of three.

\begin{figure}[t]
\includegraphics[width=0.88\columnwidth]{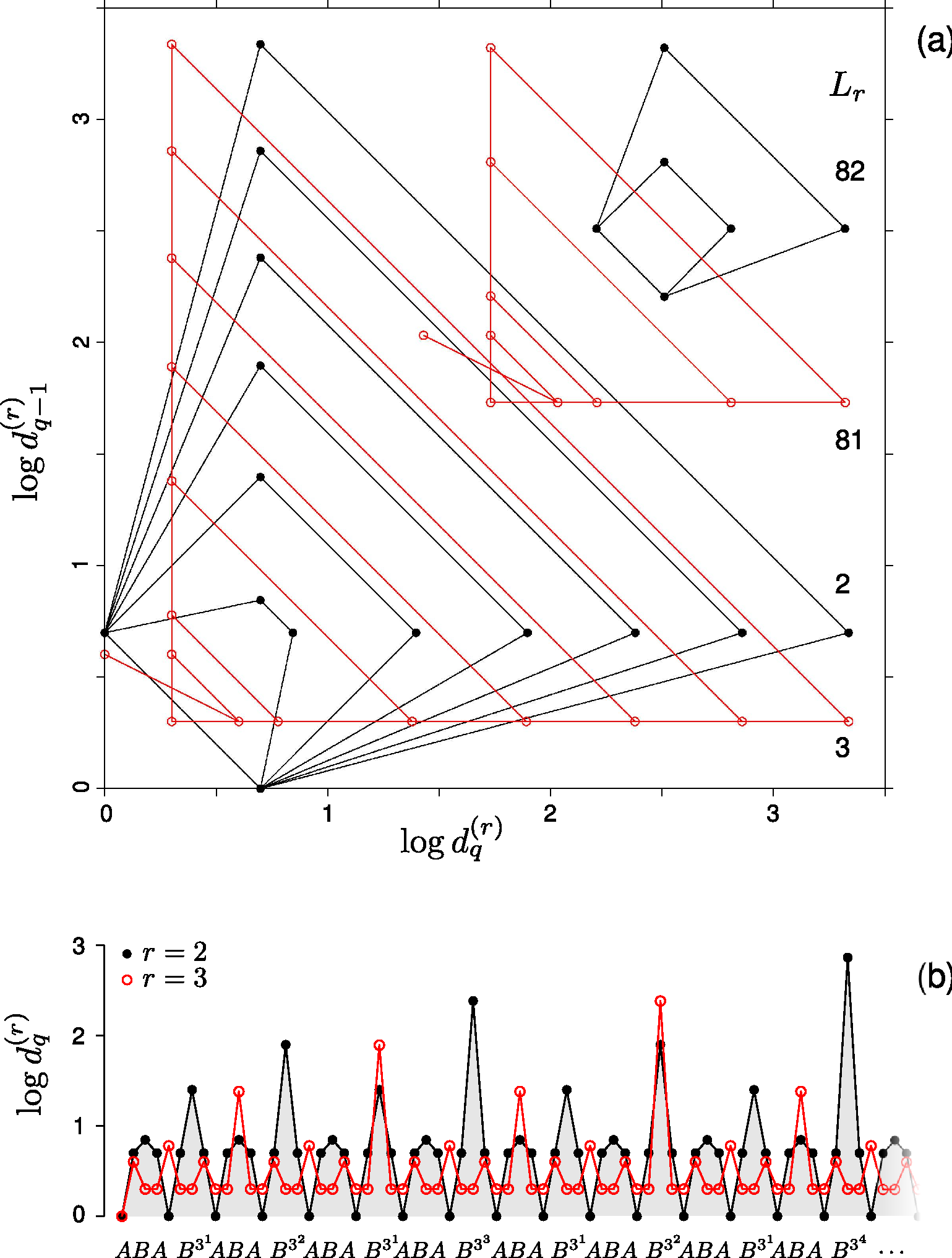}\vspace{.5cm}
\caption{\label{fig.cantor-spacing} (Color online) Local symmetry spacing (a) return maps and (b) trajectories for selected symmetry ranges in the Cantor lattice. For $L_r=r=3^k$ ($k=1,2,...$), represented here by $r = 3,81$ (and distinguished by empty circles), the spacing trajectory starts with a 'boundary anomaly' (see text).}
\end{figure}

The above features are seen in the (partial) local symmetry distribution of $w_C$ in Fig.~\ref{fig.cantor}(b).
As expected, maximal palindromes now exist for any length $l \in \mathbb Z$, as they are constructed explicitly by the inflation rule, so that $L_r = r$.
The clustering of $B$'s at increasing scale forms consecutive gaps along the $\alpha_p$-axis, since palindromes of given length are maximal within the $B$-clusters only at their borders (see upper inset in Fig.~\ref{fig.cantor}(a)).
The gaps gradually become smaller with palindrome length and close at a larger $l_p$, at which position a prefix palindrome of length $3l_p$ occurs.
Complete local symmetries thus form naturally with every iteration of the $\sigma_C$ inflation, with according lengths 
\begin{equation}L_{r=3^k} = |w^{(k)}_C| = 3^k ~~~(k = 0,1,...),\end{equation}
that is, with scaling factor $3$.

We here notice the characteristic similarity of the Cantor local symmetry distribution with that of the $w_{1,10}$ in Fig.~\ref{fig.fib101-fib110-pal-cl}(b).
In both there are $\wedge$-structures forming, with (larger) prefix palindromes occuring at the points where the legs would meet; 
however, in the $w_{1,10}$ case the gaps do not extend to the bottom, because of the absence of pure $B$-clusters (which would restrict smaller maximal palindromes to their borders).

As can be anticipated from the symmetry distribution in Fig.~\ref{fig.cantor}(b), also the spacings between maximal palindromes take on all possible values with changing $L_r$.
For fixed $L_r$, we expect the spacings to feature increasing local maxima as the symmetry axis crosses the boundaries of larger $B$-clusters deeper in the lattice, until it reaches the globally central (and largest) cluster.
In between, the spacing must return to smaller values, following the hierarchical structure of the lattice.

In Fig.~\ref{fig.cantor-spacing}(a), this behavior is shown in terms of the spacing return maps for selected local symmetry ranges.
As we see, there are no longer few-point orbits, as in the aperiodic sequences studied so far.
The number of different spacings $D^{(r)}_i$ for any single given range $L_r$ rather increases indefinitely with lattice size, i.e., with word generation $k$, and covers multiple orders of magnitude.
There is also, in general, no fixed nestedness of consecutive maps through common or collinear points.
In spite of their very different individual characteristics, however, all maps share a common asymptotic spacing scaling, given again by the Cantor partitioning number: 
\begin{equation}\lim_{i\to\infty} \frac{D^{(r)}_{i+1}}{D^{(r)}_i} = m = 3.
\end{equation}

Interestingly, there is here a 'boundary anomaly' for every $L_{3^k} = 3^k$ ($k>1$):
at these ranges, which coincide with the Cantor generation lengths and thereby with complete local symmetries, the very first axis distance is half the minimal distance for the rest of the $d^{(r)}_q$ sequence (seen as singly connected empty circles in Fig.~\ref{fig.cantor-spacing}(a)).

Each map possesses the characteristic structure originating from the underlying self-similarity in $w_C$.
This hierarchical traversal of ever larger spacings is shown in Fig.~\ref{fig.cantor-spacing}(b) for $r=2$ and $3$.
For each $L_r$, as the local symmetry axis is swept along the lattice, the spacing $d^{(r)}_q$ cycles recurrently on smaller return map orbits in hierarchical order, until the next (larger) $B$-cluster is reached, whereby a larger orbit is traversed.

Comparing the overall local symmetry properties of the Cantor lattice with those of the previously examined binary sequences, we finally note the following.
Although the Cantor lattice is generated by an inflation rule that explicitly produces palindromes (on a minimal scale and, therefore, on all scales), its local symmetry dynamics is rather 'irregular', in the sense that its spacing return maps are unbounded, have an infinite number of points, and are quite different among symmetry ranges--some even starting with the mentioned boundary anomalies.
Further, all possible ranges do occur, which is a characteristic shared statistically by a totally random lattice.
In contrast, the aperiodic lattices of Fibonacci, PD or TM type, which are generated by asymmetric inflation rules, are characterized by invariant, few-point spacing maps, with simple scaling laws along an ordered sequence of symmetry ranges.

\section{Summary and conclusion \label{conclusion}}

We have examined the local symmetry distribution (positions of locally symmetric domains as a function of their range) and dynamics (spatial evolution of the distances between local symmetry axes for given range) of one-dimensional aperiodic lattices generated by representative binary sequences, including (generalized) Fibonacci, period-doubling, Thue-Morse and Cantor.
For each case, explicit scaling behaviors and correlation properties were found numerically for the sequences of occurring maximal local symmetry ranges, and for their spacings along the lattice, in relation to the original generating inflation rules or recursive schemes.

In particular, the generalized Fibonacci lattices were shown to incorporate multiplet symmetry range subsequences with different structural features determined by the asymptotic length ratio of the corresponding binary word generations, with a dominant subsequence containing all complete local symmetries (i.e., palindromic word prefixes).
Moreover, for the Fibonacci, PD and TM lattices, the number of different local symmetry spacings is restricted to $2$ or $3$ for any symmetry range.
In terms of the subsequences of local symmetry spacing return maps for given ranges, the (asymptotic) copper Fibonacci lattice was shown to be structurally equivalent to the period-doubling lattice, of which in turn the Thue-Morse case constitutes a reduced version (with two sub-subsequences discarded).
Remarkably, the aperiodic order of each type of lattice was found to be encoded exactly in its local symmetry dynamics:
For each occurring local symmetry range, the renormalized sequence of axis spacings is generated by the inflation rule of the original aperiodic sequence.
The local symmetry distribution of the above lattices was, finally, contrasted with the Cantor lattice, whose inflation rule is inherently palindromic with gapped maximal symmetry distribution and unbounded symmetry spacing return maps.

In conclusion, we have shown that a unified view of different classes of aperiodic lattices is provided by the distribution, and, in particular, by the fixed range spacing dynamics, of their (maximal) local symmetries.
They yield a compellingly simple analysis tool to distinguish structural similarities but also lower level differences between given lattices.
The encoding of the corresponding inflation rule within any given symmetry spacing trajectory further demonstrates that local symmetries completely characterize deterministic aperiodic order at any scale.


\begin{thebibliography}{10}
\bibitem{lim}
Since we perform a numerical investigation of the considered aperiodic sequences, throughout the paper the notion of asymptotic convergence is not used in a mathematically strict sense, but refers to the observed behavior for large number of corresponding iterations.
\end{thebibliography}
\end{document}